\documentclass[]{raa}
\usepackage{deluxetable}
\usepackage{amssymb,amsmath}
\usepackage{lscape}
\usepackage{graphicx,times}
\usepackage{natbib}

\begin{document}
\def\ujy     {{$\mu Jy$}}
\def\Lya     {{Ly$\alpha$ }}
\def\spitzer {{\it Spitzer}}
\def\hst     {{\it HST}}

\title{Galaxy Formation In The Reionization Epoch As Hinted By 
Wide Field Camera 3 Observations Of The Hubble Ultra Deep Field 
}

 \volnopage{ {\bf 20xx} Vol.\ {\bf 9} No. {\bf XX}, 000--000}
   \setcounter{page}{1}

\author{Haojing Yan \inst{1}
\and Rogier A. Windhorst \inst{2}
\and Nimish P. Hathi \inst{3}
\and Seth H. Cohen \inst{2}
\and Russell E. Ryan \inst{4}
\and Robert W. O'Connell \inst{5}
\and Patrick J. McCarthy \inst{6}
}

\institute{Center for Cosmology and AstroParticle Physics, The Ohio State University, 191 West Woodruff Avenue, Columbus, OH 43210\\
  \and
   School of Earth and Space Exploration, Arizona State University, Tempe, AZ 85287
  \and
   Department of Physics \& Astronomy, University of California, Riverside, CA 92521
  \and
   Department of Physics, University of California, One Shields Avenue, Davis, CA 95616
  \and
   Astronomy Department, University of Virginia, P.O. Box 3818, Charlottesville, VA 22903
  \and
   Observatories of the Carnegie Institution of Washington, 813 Santa Barbara Street, Pasadena, CA 91101
}

\abstract{
   We present a large sample of candidate galaxies at $z\approx 7$ -- 10,
selected in the Hubble Ultra Deep Field using the new observations of the Wide
Field Camera 3 that was recently installed to Hubble Space Telescope. Our
sample is composed of 20 $z_{850}$-dropouts (four new discoveries), 
15 $Y_{105}$-dropouts (nine new discoveries) and 20 $J_{125}$-dropouts
(all new discoveries). The surface densities of the $z_{850}$-dropouts
are close to what predicted by earlier studies, however, those of the 
$Y_{105}$- and $J_{125}$-dropouts are quite unexpected. While no $Y_{105}$- or 
$J_{125}$-dropouts have been found at $AB\leq 28.0$~mag, their surface
densities seem to increase sharply at fainter levels. While some of these
candidates seem to be close to foreground galaxies and thus could possibly be
gravitationally lensed, the overall surface densities after excluding such
cases are still much higher than what would be expected if the luminosity
function does not evolve from $z\sim 7$ to 10. Motivated by
such steep increases, we tentatively propose a set of Schechter function
parameters to describe the luminosity functions at $z\approx 8$ and 10.
As compared to their counterpart at $z\approx 7$, here $L^*$ decreases by a
factor of $\sim 6.5$ and $\Phi^*$ increases by a factor of 17--90. Although
such parameters are not yet demanded by the existing observations, they
are allowed and seem to agree with the data better than other alternatives.
If these luminosity functions are still valid beyond our current detection
limit, this would imply a sudden emergence of a large number of low-luminosity
galaxies when looking back in time to $z\approx 10$, which, while seemingly
exotic, would naturally fit in the picture of the cosmic hydrogen reionization.
These early galaxies could easily account for the ionizing photon budget
required by the reionization, and they would imply that the global star
formation rate density might start from a very high value at $z\approx 10$,
rapidly reach the minimum at $z\approx 7$, and start to rise again towards
$z\approx 6$. In this scenario, the majority of the stellar mass that the
universe assembled through the reionization epoch seems still undetected by
current observations at $z\approx 6$.
\keywords{cosmology: observations --- cosmology: early universe --- galaxies: evolution --- galaxies: luminosity function, mass function --- infrared: galaxies }
}

  \authorrunning{H. Yan et al.}
  \titlerunning{Galaxy Formation in the Reionization Epoch}
  \maketitle

\section{Introduction}

  An important subject in cosmology is the hydrogen reionization of the
universe. The detections of complete Gunn-Peterson troughs
(Gunn \& Peterson 1965) in the spectra of a few $z>6$ QSOs (see Fan et al. 2006)
from the Sloan Digital Sky Survey indicate that the reionization must have ended
at around $z\approx 6$, while the recent measurement of Thomson scattering
optical depth from the seven-year Wilkinson Microwave Anisotropy Probe (WMAP)
data shows that the reionization most likely began at $z=10.4\pm1.2$ if
assuming an instantaneous reionization history (Komatsu et al. 2010). 
An important and yet controversial question is the sources of reionization. It
is clear that the QSO population can only produce a small fraction of the 
necessary ionizing photons at $z\approx 6$ (e.g., Fan et al. 2002), which
leaves star-forming galaxies as the most obvious alternative, although the
measurement of the star-formation rate density (and hence the ionizing photon
budget) is still uncertain (e.g., Stanway et al. 2003; Bouwens et al. 2003;
Giavalisco et al. 2004b; Bunker et al. 2004; Yan \& Windhorst 2004b).
Yan \& Windhorst (2004a; 
hereafter YW04a) pointed out that star-forming galaxies could account for the
entire ionizing photon budget at $z\approx 6$ as long as their luminosity
function (LF) has a steep faint-end slope, $\alpha<-1.6$. Using the Hubble
Ultra Deep Field (HUDF; Beckwith 2006) data obtained by the Advanced Camera for
Surveys (ACS), Yan \& Windhorst (2004b; hereafter YW04b) have found a large
number of $i_{775}$-dropouts, which are candidate galaxies at $z\approx 6$, and
obtained their LF that indeed has $\alpha\lesssim -1.8$. Such a very steep
slope was later confirmed by Bouwens et al. (2006) using a larger
$i_{775}$-dropout
sample collected in the HUDF, the HUDF parallel fields and the two fields of
the Great Observatories Origins Deep Survey (GOODS; Giavalisco et al. 2004a). 
Bouwens et al. (2007; hereafter B07) further compared the LF at $z\approx 4$, 5
and 6, and suggested that LF has a strong luminosity evolution over this
period in that $M^*$ is $\sim 0.7$~mag fainter at $z\approx 6$ than at 
$z\approx 4$. On the other hand, they found that the LF does not change much
in $\alpha$.  As star-forming galaxies seem to be capable of keeping the
universe ionized at $z\approx 6$,
it is natural to expect that star-forming galaxies are also capable of
producing sufficient ionizing photons at earlier epochs to make the reionization
happen, and therefore we should expect that they must
exist in significant number extending well into the reionization epoch.

   Along a different line of study, it has also been concluded that the
universe must have started actively forming stars long before $z\approx 6$.
Using the GOODS {\it Spitzer} Infrared Array Camera (IRAC) data in the HUDF
region, Yan et al. (2005) have detected the restframe optical fluxes from
three $z\approx 6$ and eleven $z\approx 5$ galaxies in the 3.6 and
4.5~$\mu$m channels, and found that these are rather matured systems with
stellar masses to the order of $\sim 10^{10}M_\odot$ and ages to the order of
a few hundred Myr (see also Eyles et al. 2005). This strongly suggests that
such high-mass galaxies must have started forming their stars at $z>7$ and
likely earlier. Yan et al. (2006) further investigate this problem using a
much larger sample in the entire GOODS field, and reinforced this conclusion
(see also Eyles et al. 2007; Stark et al. 2007). Therefore, one should indeed
expect a significant number of galaxies at $z\gtrsim 7$.

   To search for galaxies at $z\gtrsim 7$, we have to carry out surveys
in the near-IR regime, because the line-of-sight neutral hydrogen absorption 
effectively extincts the light from such sources that is emitted below 
1~$\mu m$ in observer's frame. In fact, we rely on this effect to 
identify such galaxies. The first candidate of this kind was reported by 
Dickinson et al. (2000) using the data obtained by the Near Infrared Camera
and Multi-Object Spectrometer (NICMOS) No. 3 (NIC3) in the Hubble Deep
Field, although now we have convincing evidence that it is likely a red galaxy
at lower redshift, similar to those ``IRAC-selected Extremely Red Objects''
(IERO; Yan et al. 2004). Using the two broad-band data taken by NIC3 in
the HUDF (Thompson et al. 2005), YW04b identified three objects that are
missing from the ACS images. Subsequent analysis using the Spitzer IRAC data
suggest that two of them are red galaxies at $z\approx 2$--3 without much 
on-going star formation (Yan et al. 2004; but also see Mobasher et al. 2005
and Chary et al. 2006), while the third one (ID No.3) is less conclusive
because it is blended with other sources in the IRAC image and thus its
photometry is uncertain. Bouwens et al. (2004) used the same
data set, and pushed to a fainter limit and found a few additional candidates. 
Bouwens et al. (2005) further included all available deep NIC3 imaging data
and extended their search to $z\approx 10$, although no conclusive answer
was obtained. Using these results, Bouwens et al. (2008; hereafter B08) derived
constraints to the UV LF of galaxies at $z\approx 7$--10, and argued that
the LF evolves strongly and continues the dimming trend in $L^*$ (as proposed in
B06) to higher redshifts. As a result, the number density of galaxies, and
hence the UV luminosity density in the earlier universe should be significantly
lower than at a later time. 

   Meanwhile, the search for gravitationally lensed galaxies around foreground
galaxy clusters have resulted in some remarkable success. Kneib et al. (2004)
first discovered such an object that is likely at $z>6$. While no precise
redshift was obtained, extensive optical and IR observations suggest that it
is at $z\approx 7$.
Bradley et al.  (2008) reported a very bright, highly probable candidate at 
$z\approx 7.6$. Zheng et al. (2009) reported three new $z\approx 7$ candidates
from the same campaign. 
Richard et al. (2008) have found 12 candidates at similar redshifts. 
These results, however, still do not allow us put a
strong constraint to the number density of galaxies at these redshifts, although
they indeed prove that very high-redshift galaxies much fainter than our
current detection limits (should there be no lensing magnification) do exist.
There are also some other evidence from the search for \Lya emitters around 
foreground clusters that supports similar conclusion (Stark et al. 2007).

  While the above results start to give us some meaningful constraints
to the faint-end of the LF at $z\gtrsim 7$, there is still only limited
constraint to the bright-end where deep wide-field surveys are required. 
To date, most such surveys have null detection (e.g., Willis et al. 2008;
Stanway et al. 2008; Sobral et al. 2009) or are uncertain (e.g. Hickey et al.
2009). One exception that has produced positive detections is the Subaru
CCD survey for LAEs at $z=7.0\pm0.1$, which has resulted in the redshift
record of $z=6.96$ (Iye et al. 2006). However, even such LAEs can only
constrain the faint population, as the vast majority of them are pure
emission-line objects and are not seen in continuum. Another exception is
that of Ouchi et al. (2009), which surveyed $\sim 1500$~arcmin$^2$ in
two fields and resulted in 22 $z\approx 7$ candidates to $AB\sim 26.0$~mag.
Recently, Capak et al. (2009) reported three very bright candidates at
$z\approx 7$ at $J\sim 23$~mag, however, their nature is still uncertain.

   Nearly all these studies (with the exception of Capak et al. 2009) in the
very high-redshift frontier have claimed that
their results are consistent with the strong declining evolution of the LF 
with respect to increasing redshift as suggested by B07 and B08. If this is
indeed what the universe behaves, we are facing a dilemma: on one hand, the 
hydrogen reionization, which is now well constrained that must have started at
$z\approx 10$, requires a large amount
of strong UV emitting sources, and the significant stellar mass density
measured at $z\approx 6$ also strongly suggests very active star formation
activities at $z\gtrsim 8$ and above; on the other hand, the limited number
of observations to search for galaxies beyond $z\approx 7$ indicates a strong
declining number density of galaxies at higher redshifts. To reconcile these
seemingly conflict results, a more decisive survey for galaxies at
$z\gtrsim 7$ is in demand.

   The Wide Field Camera 3 (WFC3) recently installed to \hst\, has provided
a unique opportunity for the study of the universe at very high redshifts. The
IR channel of this camera has a factor of $6.4\times$ larger field-of-view
(FOV) and an order of magnitude higher Q.E. as compared to NIC3, making it the
most powerful tool in detecting galaxies at $z\approx 7$ and beyond. For this
reason, the first set of deep data that it took has already inspired four
papers to appear at the arXiv preprint service within one week after the data
were made public (Bouwens et al. 2009; Oesch et al. 2009; Bunker et al. 2009;
McLure et al. 2009). All these new results, however, seem to reiterate that 
the number density of galaxies rapidly declines when we look back further in
time. Here we present our results based on an independent reduction and 
analysis of these data. Our effort has resulted in more candidate galaxies
at $z\gtrsim 7$ than others, and, for the first time, a large sample of
highly probable candidate galaxies at $z\approx 10$.
We will show that our analysis has led to a completely new, although still
tentative conclusion about the formation and evolution of galaxies in the
early universe.

   Our paper is organized as following. We briefly describe the WFC3 IR
instrument and its observations of the HUDF in \S 2, and give the details of
our data reduction in \S 3. Our photometry and catalog construction is
described in \S 4. The candidate selection and the dropout samples are
presented in \S 5. We discuss the implications of our results in \S 6, followed
by a summary in \S 7. For simplicity, we denote the ACS passbands F435W, F606W,
F775W, and F850LP as $B_{435}$, $V_{606}$, $i_{775}$, and $z_{850}$,
respectively, and denote the three WFC3 IR passbands F105W, F125W and F160W as
$Y_{105}$, $J_{125}$ and $H_{160}$, respectively. All magnitudes quoted are in
AB system. Throughout the paper, we use the following cosmological
parameters: $\Omega_M=0.27$, $\Omega_\Lambda=0.73$ and 
$H_0=71$~km~s$^{-1}$~Mpc$^{-1}$.

\section{WFC3 IR Observations of the HUDF}

   We have used the first epoch of data from the HST Cycle-17 General Observer 
Program of Illingworth et al. (PID. GO-11563; Bouwens et al. 2009; Oesch et al.
2009), which was designed to do very deep WFC3/IR observations in three
existing, deepest HST fields.
Below we briefly describe the instrument and the observations.

   The detector of WFC3/IR is a HgCdTe array with 1024$\times$1024 square pixels
of $18\mu m\times 18\mu m$ in size, and is sensitive from 400 to 1700~nm
(HST WFC3 Handbook; Kimble et al. 2008).
To complement WFC3/UVIS, WFC3/IR is equipped with filters at wavelengths
$\gtrsim$~800~nm, and its performance is optimized for wavelengths at
$\gtrsim$~1000~nm, where the detector sensitivity reaches its peak. It uses CMOS
circuits to make non-destructive readouts in four separate $512\times 512$
quadrants. The effective total area of the array is $1014\times 1014$ pixels,
as the outer 5~pixels around the four edges of the detector are blocked from
light and are used as ``reference pixels'', which provide constant-voltage
reference values for the readout circuits to help monitor and remove the drift
in the electronics. For this reason, WFC3/IR does not suffer from the
``pedestal'' variations that are common to NICMOS data. While it replaces WFPC2
(and therefore is on-axis), WFC3 still has significant geometric distortion
because its focal plane is tilted with respect to the optical axis. The 
projected IR pixels on the sky are rectangles with scales of 
$0.121^{''}\times0.135^{''}$ and the camera has a total FOV of 
4.67~arcmin$^2$. The camera can be operated in either the imaging mode or the
slitless spectroscopy mode.

    Program GO-11563 is to do very deep WFC3/IR imaging in the HUDF
(Beckwith et al. 2006) and the two HUDF05 fields (Oesch et al. 2007), taking
advantage of the extremely deep ACS optical imaging data that already exist.
It uses three broad-band filters, namely, $Y_{105}$, $J_{125}$ and $H_{160}$.
As illustrated in Fig. 1., this filter combination offers the
capability of selecting galaxies at $z\gtrsim 7$--10 using the dropout technique
to detect the Lyman-break signature that occurs at restframe 1216\AA.
This program has been allocated 192 orbits: 22 orbits of $Y_{105}$, 36 orbits of
$J_{125}$ and 38 orbits of $H_{160}$ are for the HUDF, while 11 orbits of 
$Y_{105}$, 18 orbits of $J_{125}$ and 19 orbits of $H_{160}$ are for each of
the HUDF05 fields. The
intended depths are $\sim 29.0$ mag in the HUDF and $\sim 28.6$ mag in the
HUDF05. The observations are packed in visits of four orbits, with each orbit
being split into two 1402.9-second exposures operated at the MULTIACCUM mode
using the SPARS100 sequence of 15 samplings. A ``half-pixel'' dithering
strategy is used to improve image quality and resolution.

\begin{figure}[tbp]
\centering
\includegraphics[width=9.0cm]{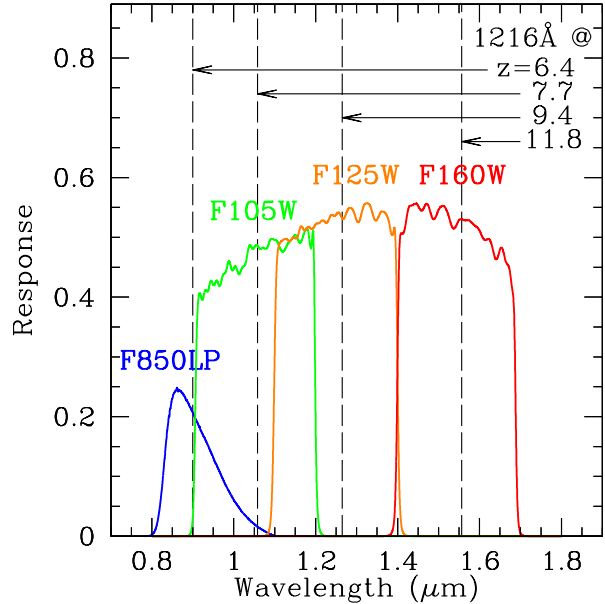}
\caption{System response curves of the four major passbands used in this study.
Dashed vertical lines indicate the locations of the Lyman-break signature
(which occurs at the restframe $\sim$~1216\AA\, for galaxies at $z\gtrsim 4$)
at various redshifts and demonstrates how galaxies at progressively higher
redshifts can be selected as dropouts through successive passbands moving to
the redder wavelengths.
}
\end{figure}

   The first epoch of observations were done in the HUDF only, and the data were
taken from August 26, 2009 to September 6, 2009. The remaining observations
will be executed in year 2010. The obtained data, which include 18 orbits
in $Y_{105}$, 16 orbits in $J_{125}$ and 28 orbits in $H_{160}$, were made
public on September 9, 2009. The observations were oriented to have a similar
position angle as the HUDF ACS images, and the field was centered on
RA=3$^{h}$32$^{m}$38.5$^{s}$, DEC=$-$27$^{\rm o}$47$^{'}$0.0$^{''}$ (J2000.0).
As mentioned in Bouwens et al. (2009) and Oesch et al. (2009), two orbits in
$Y_{105}$ were affected by image persistence from an
earlier exposures in an unrelated program, and these data were excluded from
their analysis. Bunker et al. (2009) and McLure et al. (2009) opted to do the
same. In fact, another two orbits in $H_{160}$ were also
affected by image persistence to the same degree. However, the serious
effects are limited to a small region close to the center of the detector. We
have decided to include these images in our analysis. As described below, our
reduction has successfully masked out the affected pixels, and thus we have
taken full advantage of the entire data set. The total effective exposure
times are 50505.7, 44894.0, and 78564.5 seconds in $Y_{105}$, $J_{125}$ and
$H_{160}$, respectively. 

\section{Data Reduction}

   In this section we provide the details of our data reduction procedure.

\subsection{Basic Calibration}

   The public data released on September 9, 2009 include the raw images and the
\hst\, pipeline processed images. While the latter ones are the flat-field
corrected products, their Data Quality (DQ) extensions have been populated
with masks of cosmic rays identified by the pipeline using default parameters
that were still preliminary. Therefore, we re-processed the raw
images on our own, using the CALWF3 task (Version 1.6, implemented on April 27,
2009) included in the STSDAS package (Version 3.10) and the latest reference
files (as of September 9, 2009). Our first-step products from this procedure
are the same as those distributed in the public archive, with the exception
that the DQ extensions only include the ``static mask'' that identifies the bad
pixels of the detector.

   During this step, we noticed that the four quadrants of the processed
images had notable offsets at the level of 1--3\%. The inspection of the
processed in the public archive revealed the same problem. This seems to
suggest that there are quadrant-dependent bias residuals similar to the
so-called ``pedestal'' in a NICMOS image. However,
this was unexpected for the WFC3/IR data, as the reference pixels have been
used to remove such drifts (see \S 2). It was soon discovered that these
offsets were actually caused by a confusion in assigning calibration steps,
which has led CALWF3 to apply the gain corrections of the four quadrants twice
\footnote{This problem was later reported in the September 2009 issue of the
WFC3 Space Telescope Analysis Newsletter (STAN); see 
http://www.stsci.edu/hst/wfc3/documents/newsletters/STAN\_09\_01\_2009.
This problem was fixed in the latest CALWF3 (Version 1.7).}.
We manually fixed this problem by applying the appropriate corrections
(H. Bushouse, priv. comm.) to
the four quadrants. The end result is very satisfactory; the offsets disappear
completely
\footnote{We note that a multiplicative correction of this kind was not made in
Bouwens et al. (2009), Oesch et al. (2009) or Bunker et al. (2009), 
and this may have contributed to the differences in photometry.
McLure et al (2009) treated the four quadrants independently and thus avoided
this problem (A. Koekemoer, priv. comm.).
}.
We further inspected the sky-subtracted, median-combined images in each band,
and noticed that a small, smooth background gradient was still present at 
the $\sim$ 1\% level. This
gradient was removed by fitting a plane to each individual image; the mean
sky was also subtracted in this step.

\begin{figure}[tbp]
\centering
\includegraphics[width=9.0cm]{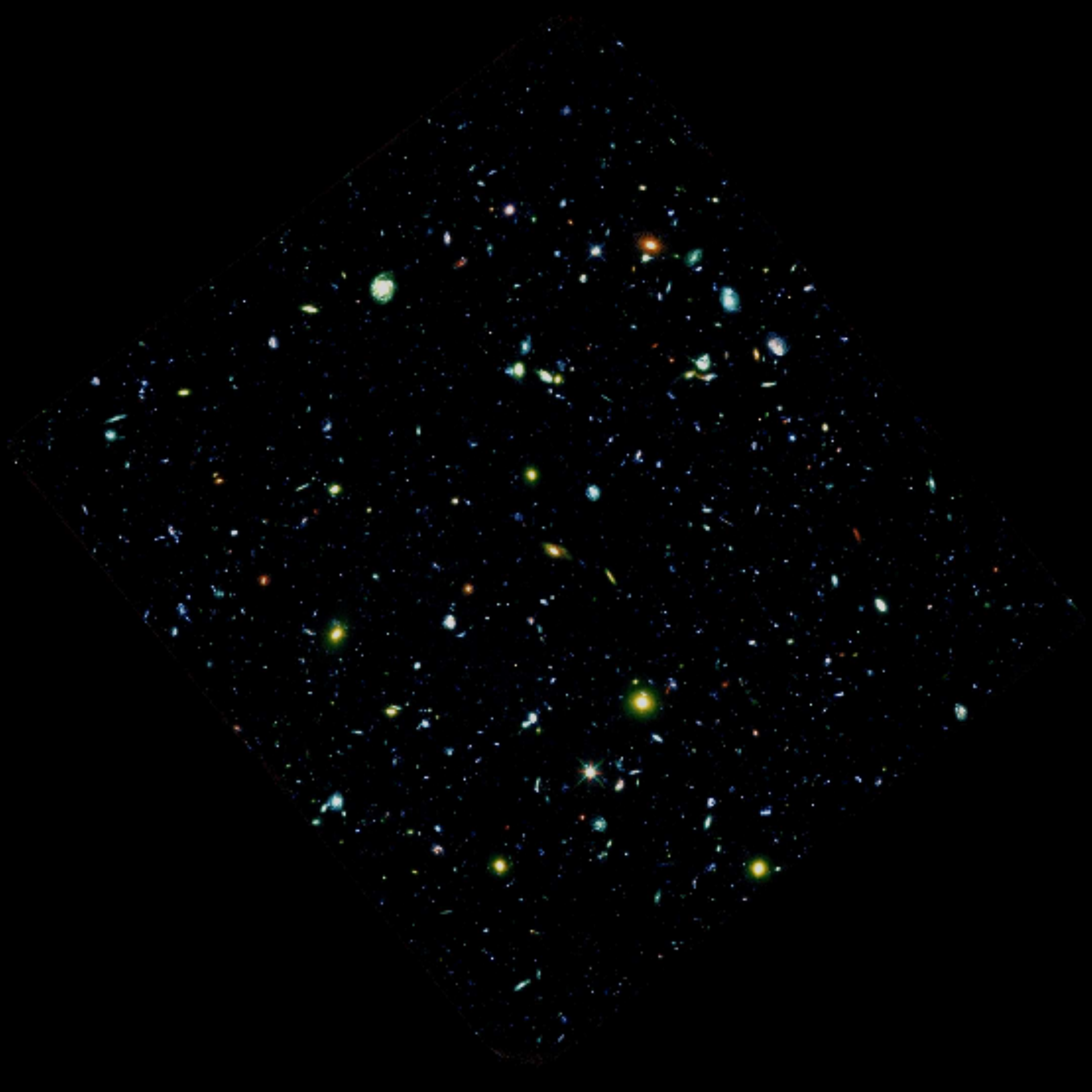}
\caption{Color-composite image of the HUDF, created using
the weighted-mean of the three-band WFC3/IR mosaics that we made as red,
the ACS $i_{775}$ as green, and the ACS $B_{435}$ as blue. [Due to file
size limit only a low-resolution picture is displayed here.]
}  
\end{figure}

\subsection{Image Mosaicking}

   We used the MultiDrizzle software (Koekemoer et al. 2002) distributed in the
STSDAS.DITHER package (Version 2.3; May 11, 2009) to correct for geometric
distortion and to stack the individual images. MultiDrizzle is a Python wrapper
that streamlines the processing of \hst\, images using the Drizzle algorithm of
Fruchter \& Hook (1997) to optimally recover spatial resolution by subpixel
resampling. While its implementation for handling the new WFC3 data is
understandably still preliminary at this stage, it has served our main purpose
well, although we had to make quite a number of manual adjustments.

   A critical part of this step is to correct for geometric distortion, which
affects essentially everything including cosmic-ray rejection, astrometric
accuracy and photometry. We used the latest distortion parameter table
(IDCTAB; made public on September 9, 2009) delivered by the WFC3 instrument
team, which were derived based on the in-flight calibration observation on
July 23, 2009.  While this calibration was done using $H_{160}$ only, the
polynomial representations are the same for all bands and thus is taken as
universally applicable in the IDCTAB. We note that McLure et al. (2009) used
the same IDCTAB as we did, while Oesch et al. (2009), Bouwens et al. (2009) and
Bunker et al. (2009) derived their own geometric distortion solution using
the HUDF ACS data as the reference.

   As usual, images obtained during different visits still have non-negligible
offsets in their World Coordinate System (WCS) after correction for the
geometric distortion. This is mainly caused by the intrinsic
astrometric inaccuracies of the guide stars used in different visits. If this
were not corrected, MultiDrizzle would not properly identify the cosmic-ray
hits, or would incorrectly flag real objects as cosmic rays. More critically,
the individual images would not be aligned and thus would not result in a useful
stack. To solve this problem, we used the HUDF ACS images as the astrometric
reference and calculated the transformation that should be applied to each
{\it distortion-corrected} image. About 11--12 common objects were manually 
identified in each input image and the reference ACS $z_{850}$ image, and we
solved for X-Y shift, rotation, and plate scale. Here the ACS image was
$3\times 3$ rebinned, giving a spatial resolution of 0.09$^{''}$/pix. This scale
would be the plate scale of our final WFC3/IR stacks, and was chosen for a
good reason. We would like to obtain a better spatial resolution offered by the
subpixel-dithered observations, and at the same time we do not want to lose
sensitivity to low surface brightness objects because of excessive
subsampling. The chosen scale is a well matched compromise of all factors
considered. 

   These transformations were then passed to MultiDrizzle, and the drizzling
process was re-run to put each input image onto the pre-specified grid. We
set the drizzling scale (``pixfrac") to 0.8. Cosmic-ray hits were identified
and excluded from stacking. We inspected the few images that were affected by
the image persistence, and verified that the affected pixels were flagged in
the masks and did not enter the combining process. The end 
results were mosaics and associated weight images in three bands, all aligned
with the $3\times 3$ rebinned ACS images. By comparing the positions of
point-like sources in the ACS images and in our mosaics, we conclude that the
astrometry of the latter is good to $\sim 0.5$~pixel (0.045$^{''}$) on
average; we expect that this can be further improved after
better geometric distortion coefficients are available. Fig. 2 shows a color
composite of the field, using the ACS $B_{435}$ as blue, ACS $i_{775}$ as
green, and a weighted mean of all three WFC3/IR bands as red.

   We note that McLure et al. (2009) also used the standard MultiDrizzle to
generate mosaics, and they adopted a final scale of 0.03$^{''}$/pix. 
Oesch et al. (2009) and Bouwens et al. (2009) used their own version of
modified MultiDrizzle, and adopted a final scale of 0.06$^{''}$/pix.
Bunker et al. (2009) did not use the drizzle algorithm but combined the
individual images after using IRAF GEOMAP and GEOTRAN tasks to map them to the
same astrometric grid of a 0.06$^{''}$/pix scale.

\section{Catalog Construction}

  The selection of dropouts relies on comparing photometry of sources
in successive passbands. Our photometry procedure, which is optimized for this
purpose, is described below in detail.

\subsection{Photometry}

   We carried out matched-aperture photometry by running the SExtractor program
of Bertin \& Arnouts (1996) in dual-image mode. The three WFC3/IR mosaics were
used in turn as the detection images to prepare catalogs for dropout selection
at different redshifts. Hereafter we will refer to these catalogs as the
$Y_{105}$-based, $J_{125}$-based and $H_{160}$-based catalogs, respectively.
Our photometric zeropoints were from the latest delivery of the WFC3
instrument team on September 9, 2009; in AB system, these are 26.27, 26.25 and
25.96 mag in $Y_{105}$, $J_{125}$ and $H_{160}$, respectively
\footnote{See http://www.stsci.edu/hst/wfc3/phot\_zp\_lbn, and also the WFC3
STAN September 2009 issue.}.

   The drizzle-combined images have correlated pixel noise because of the
subpixel sampling. This correlation should be taken into account when
estimating photometric errors of measured sources. We followed the procedure
utilized in the GOODS program (Dickinson et al. 2004) to calculate the
correlation amplitudes in our mosaics, and then to derive the so-called 
``RMS maps" that describe the absolute root-mean-square noise of each pixel.
We found that the actual noise would be underestimated in the WFC3 mosaics by
a factor of 1.2--1.7 if the correlation were not properly taken into account.
Similar procedure was applied to the
the rebinned ACS images as well, because the rebinning also changed the pixel
noise properties. The resulting RMS maps were supplied to SExtractor to
estimate background fluctuation at the source detection phase and to calculate
photometric errors at the extraction phase. We have also estimated the depths
of the WFC3/IR mosaics using these RMS maps. The 5-$\sigma$ depths within a
0.2$^{''}$-radius aperture are 28.52, 28.85 and 28.94 mag in $Y_{105}$,
$J_{125}$ and $H_{160}$, respectively. We note that these values are generally
consistent with the depths quoted in other works. However, they do not
completely agree --- our estimates tend to be slightly shallower. 
The differences could
be due to two major reasons. One reason is the slight difference in zeropoints
used, which we will discuss further in next section. The other is the 
differences in the methods used to measure the correlated noise; 
underestimating the correlation noise would result in artificially deeper
limits and artificially higher signal-to-noise ratio (S/N).

   The SExtractor runs were set to output two different flavors of magnitudes,
namely, MAG\_AUTO and MAG\_ISO. The default SExtractor parameter settings for
MAG\_AUTO use a Kron factor of 2.5 and a minimum radius of 3.0~pixels, and the
resulting MAG\_AUTO values are generally considered as total magnitudes. As we 
are mostly interested in faint objects for this study, we set these parameters
to (Kron factor, minimum radius)$=$(1.2, 2.0). This results in small apertures
for MAG\_AUTO measurement, which, while losing a non-negligible amount of
light for bright objects, is optimal for high S/N
extraction of faint sources. With these settings, the MAG\_AUTO magnitudes are
quite similar to MAG\_ISO, except that at the very faint levels the latter
usually result in better S/N as their extraction apertures tend to be smaller.
We compared the MAG\_AUTO values with these two different settings for the
sources that are brighter than 22.0 mag, and found a systematic offset of
0.16 mag. We thus add $-0.16$ mag to correct the MAG\_AUTO values obtained with
small Kron parameters to total magnitudes. 

   A 2-pixel, $5\times 5$ Gaussian filter was used in these SExtractor runs 
for source detection. The detection threshold was set to 0.8-$\sigma$ above the
filter-convolved image, and a minimum of 2 connected
pixels above the threshold was required. For an extracted source to be included
in the final catalog, it is required to have $S/N\geq 3$ in the detection band,
calculated within either the MAG\_AUTO aperture {\it or} the MAG\_ISO aperture.
Excluding the noisy field edges, our final $Y_{105}$-based, $J_{125}$-based and
$H_{160}$-based catalogs include 2475, 2945 and 3156 sources, respectively.
In contrast, Bunker et al. (2009) quoted much more conservative extraction
criteria and yet reported $\sim 4000$ sources in their $Y_{105}$-based catalog.

   Fig. 3 presents the source surface densities in these three bands inferred
from our catalogs. For comparison, the source surface density derived from the 
$z_{850}$-based catalog of YW04b is also shown. All these curves agree with
each other very well up to AB$=$29.2 mag (indicated by the vertical dotted 
line). Beyond this point, the $Y_{105}$ count drops slightly faster than the
$z_{850}$ count, while the $J_{125}$ and $H_{160}$ counts appear to be much
higher than the $z_{850}$ count, which seems to suggest that source
contamination becomes a problem. Therefore, we limit the candidate search to 
$AB\leq 29.2$~mag. When interpreting our results later in the paper, we
further limit our discussion to $AB\leq 29.0$~mag (see \S 6).

\begin{figure}[tbp]
\centering
\includegraphics[width=9.0cm]{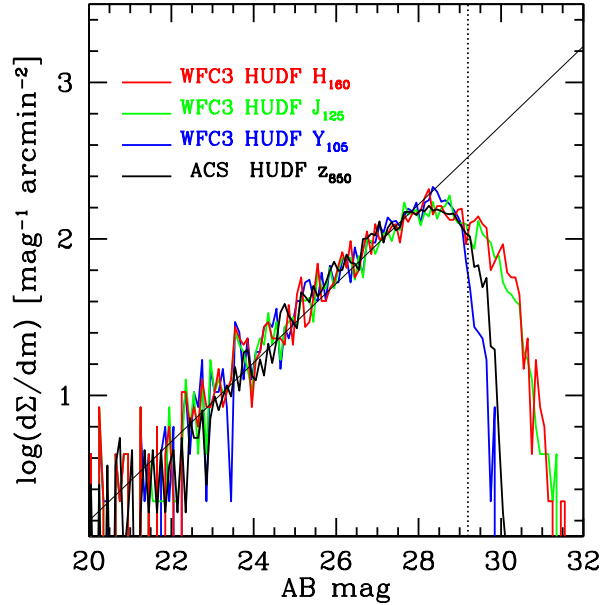}
\caption{Source surface densities in the three WFC3 passbands, derived using
the source counts from the $Y_{105}$-, $J_{125}$- and $H_{160}$-based 
catalogs, respectively. For comparison, the black curve shows the source
surface density in the ACS $z_{850}$-band, derived using the HUDF
$z_{850}$-based from YW04b. In all cases, only sources with 
$S/N\geq 3$ are used. The vertical dotted black line is at AB=29.2 mag,
beyond which the $J_{125}$- and $H_{160}$-based catalogs seem to start to
suffer from contamination. We limit our analysis to the regime brighter than
this limit. For reference, the thin straight line indecates a power-law fit
to the part where the counts are complete.
}
\end{figure}

\subsection{Zeropoints}

   While Oesch et al. (2009) and Bouwens et al. (2009) have used the same WFC3
magnitude zeropoints as we did, Bunker et al. (2009) and McLure et al. (2009)
have used different methods to derive their own zeropoints. These different
approaches all involve using either the NIC3 or the VLT/ISAAC observations in
this area. In this section we examine to what extend these existing
observations are consistent with the new WFC3 data in terms of zeropoints.

   The ISAAC mosaics and their zeropoints used here are from the ESO
GOODS/ISAAC final data release (Version 2.0) as of September 10, 2007
(Retzlaff et al. in prep.), which were based on the data obtained under
Programs LP168.A-0485. We only compared to the ISAAC J-band, as this passband
is close to WFC F125W-band (see the bottom-left panel of Fig. 4). We used
MAG\_AUTO, and the photometry of the ISAAC image was done using
(Kron factor, minimum radius)$=$(2.5, 1.2) and hence the obtained MAG\_AUTO can
be treated as total magnitudes. As mentioned earlier, we already added $-0.16$
mag to our WFC3 photometry to correct to total magnitudes. The comparison of
WFC3(125W) vs. ISAAC(J) is shown in the top-left panel of Fig. 4. The average
offset is $\sim 0.04$~mag, suggesting that a proper zeropoint can be obtained in
this band by using the ISAAC J-band as reference.

  The NIC3 mosaics used here are the Version 2 products of Thompson
et al. (2005), which are registered to exactly the same astrometric grid and
have the same pixel scale as ours (i.e., using the $3\times 3$ rebinned HUDF
ACS images for reference). We only compared our F160W image to their F160W one,
as the two passbands are reasonable close to each other. We did not compare to
their F110W band, as the latter is much wider than either the WFC3 F105W or
F125W and thus the a straightforward comparison would be difficult.
We run SExtractor in dual-image mode with the same parameter settings as
described above, using their NIC3 mosaic as the detection image. The
comparison between WFC3(F160W) and NIC3(F160W) in the right panel of Fig. 4. 
The offset between the two sets amounts to $\sim 0.15$~mag (with NIC3
magnitudes being too faint). Similar results have been reported by Coe et al.
(2006), who quote a zeropoint offset of $0.18$~mag in NIC3(F160W). In addition,
a significant trend of non-linearity in the NIC3 magnitudes can be clearly seen
in Fig. 4. This non-linearity was not
determined until 2006 (see NICMOS ISR 2006-001 \& 002), and therefore was not
removed from the NIC3 HUDF data. Cautions should be
taken when using the NIC3 images for zeropoint determination.

\begin{figure}[tbp]
\centering
\includegraphics[width=9.0cm]{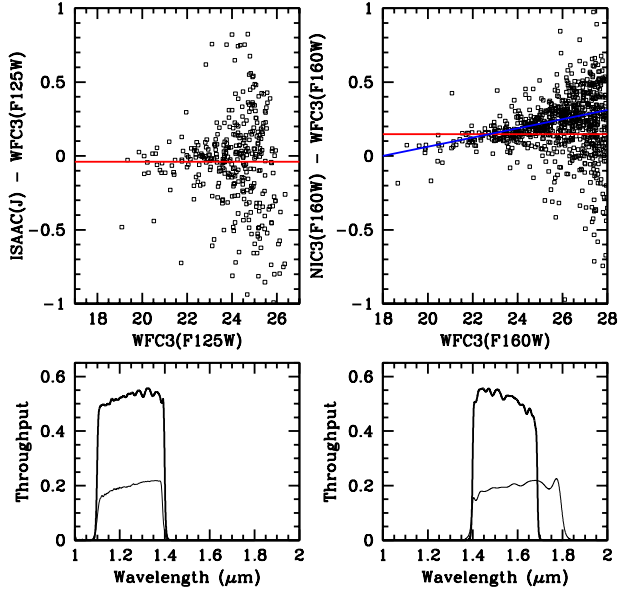}
\caption{Magnitude zeropoint comparisons to those using the VLT/ISAAC images
(left) or the NIC3 images (right) as references. Only comparisons to the ISAAC
J-band and NIC3 F160W-band are made, as the shapes of these two passbands 
(thin lines in the bottom panels) are reasonably close to WFC3 F125W and F160W
(thick lines in the bottom panels), respectively. 
While the ISAAC J-band magnitudes
appear to be $-0.04$~mag too bright on average (indicated by the horizontal
line in the top left panel), the NIC3 F160W-band magnitudes are $+0.15$~mag
too faint (indicated by the horizontal line in the top right panel) if using
the zeropoint of Thompson et al. (2005). In addition, the offset of NIC3
F160W-band magnitudes show a general trend with respect to source
brightness (as shown by the blue line), indicative of a significant
non-linearity in NIC3 zeropoint calibration.
}
\end{figure}

\subsection{Photometry of Other Groups}
 
   As different approaches to photometry can result in different catalogs and
thus impact the candidate selection, for comparison purpose we summarize here
the methods and some important settings that were used in the other four papers.


   Bunker et al. used 
the thermal-test zeropoints of 26.16, 26.10, 25.81 mag in these three bands,
which are 0.11, 0.15 and 0.15 mag too bright, respectively. McLure et al.
derived their own zeropoints using ground-based near-IR observations and also
the NICMOS F160W-band image, and obtained zeropoints of 26.25, 26.25 and 26.10
in $Y_{105}$, $J_{125}$ and $H_{160}$, respectively. These are very close to
the in-flight zeropoints that we used; the offsets are $-0.02$, 0.00 and 0.04
mag in these three bands, respectively.

   All groups except McLure et al. have used SExtractor in dual-image mode for
both the rebinned ACS and the WFC3 IR images. Bouwens et al. used the
$\chi^2$-combined $J_{125}$ and $H_{160}$ image as the detection image.
They used a small Kron factor of 1.2 as we did, but corrected these magnitudes
to total magnitudes on a source-by-source basis. Finally, they added an overall
correction of $-0.1$~mag to account for light on the wings of the PSF. Oesch
et al. used the summed $Y_{105}$ and $J_{125}$ image as the detection image,
and used Kron factor of 2.5 to define the apertures. They also added a final
correction of $-0.1$~mag. For comparison, we did not apply this $-0.1$~mag
correction. Bunker et al. used their $Y_{105}$, $J_{125}$
and $H_{160}$ mosaics as the detection images for the $z_{850}$-, $Y_{105}$-
and $J_{125}$-dropout selections, respectively, similar to what we did. They
used a fixed aperture of $r=0.3^{''}$ for flux measurement. McLure et al.
performed photometry on the three bands independently, and used a fixed
aperture of $r=0.3^{''}$.

   All these subtle differences, in addition to the differences in the data
reduction, contribute to the differences in the photometric results of
different groups, especially at the very faint levels.

\section{Dropout Selection and Samples}

   We have mainly used the $Y_{105}$-, $J_{125}$- and $H_{160}$-based catalogs
to select $z_{850}$-, $Y_{105}$- and $J_{125}$-dropouts as candidate galaxies
at $z\approx 7$, 8 and 10, respectively. We have used a ``$J_{125}$+$H_{160}$''
image to improve the completeness of the $Y_{105}$-dropouts. This section
presents the details of the selection and the 
dropout samples. While our catalogs include both MAG\_AUTO and MAG\_ISO, the
latter is only for quality control purpose;
MAG\_AUTO is used throughout the remainder of this paper.

\subsection{Selection of $z_{850}$-dropouts}

   Our major color criterion for $z_{850}$-dropout selection is
$z_{850}-Y_{105}\geq 0.8$~mag, which select candidates at $6.4\leq z\leq 7.7$.
We synthesized a large number of galaxy templates based on the model of 
Bruzual \& Charlot (2003), and found that one should use the criterion of
$z_{850}-Y_{105}\geq 1.0$ mag to select galaxies at $z\gtrsim 6.6$. Lowering
this threshold to $z_{850}-Y_{105}\geq 0.8$ mag is appropriate for including
galaxies at $z\approx 6.5$. As the $i_{775}$-dropout selection for galaxies at
$5.5\lesssim z\lesssim 6.5$ loses its efficiency at $z\approx 6.5$ where
the Lyman-break signature is starting to move out of $z_{850}$-band, we decided
to adopt $z_{850}-Y_{105}\geq 0.8$~mag in order to obtain
a more complete census of high-redshift galaxies. 
In addition to the above criteria, we required that a candidate should be
undetected ($S/N<2$) in the other ACS bands 
($i_{775}$, $V_{606}$ and $B_{435}$). 
The color-color diagram shown in Fig. 5 demonstrates our selection.
We did not place constraint on $Y_{105}-J_{125}$ color, as it depends on
the intrinsic properties of the galaxy that we do not know {\it a priori}.
While this could potentially result in slightly more contamination from very
cool Galactic brown dwarfs and some very red galaxies at lower redshifts,
our procedure below largely eliminates this concern.

\begin{figure}
\centering
\includegraphics[width=9.0cm]{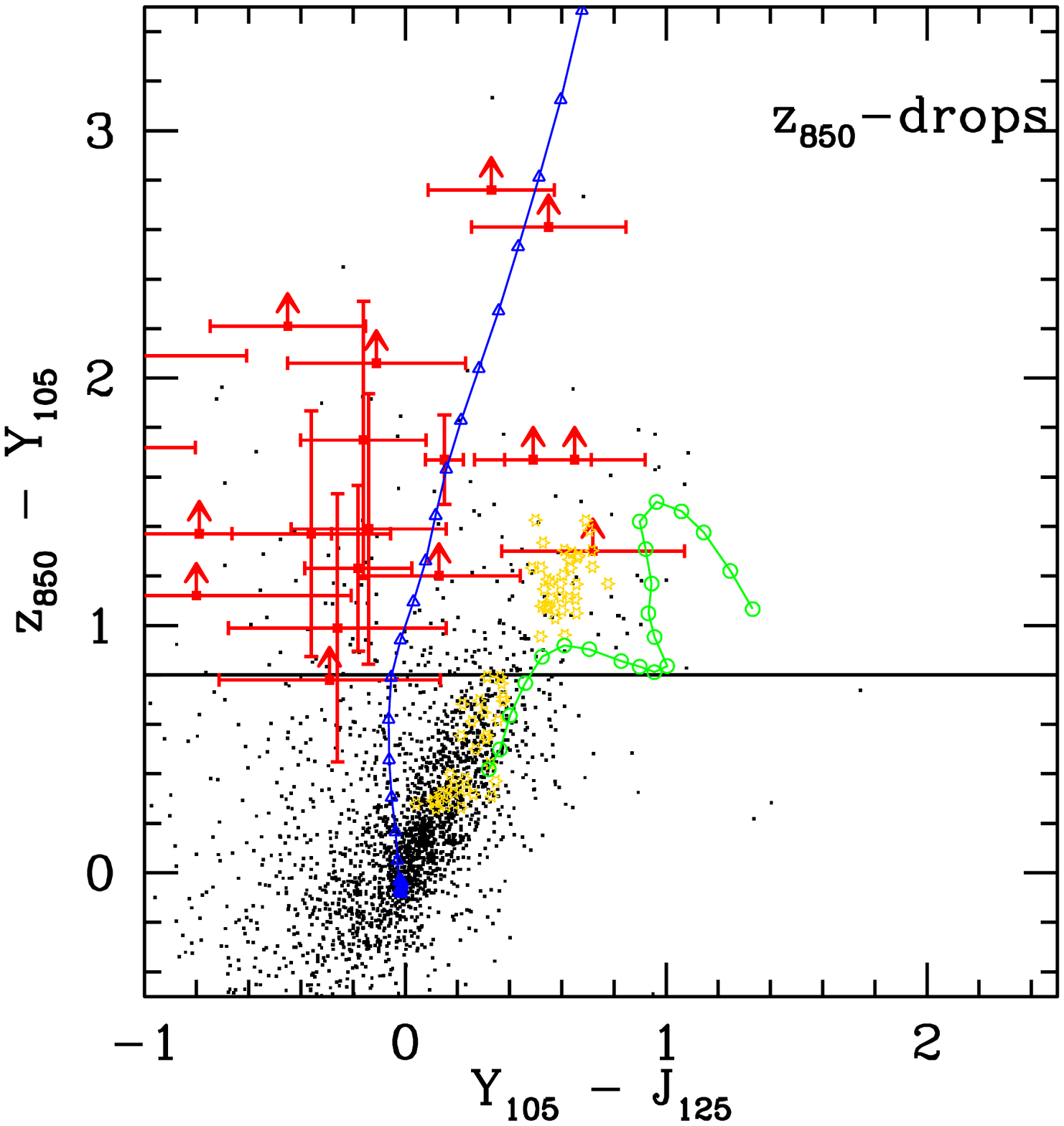}
\caption{Color-color diagram demonstrating the selection of $z_{850}$-dropouts.
The black dots are field objects, and the horizontal line indicates the
major color selection criterion of $z_{850}-Y_{105}\geq 0.8$~mag. The red
data points with error bars are the selected candidates. The yellow symbols
show the colors of M-, L- and T-dwarfs, which are calculated using the
spectra from (Leggett et al. 2000 ApJ 535 965), while the green symbols
show the colors of a typical red galaxy at $z\approx 1$--3 simulated using BC03
models. The blue symbols shows the color track of a typical young galaxy at
high redshifts; at $z\gtrsim 6.5$ it passes above the selection threshold.
}
\end{figure}

   The WFC3 and ACS images of the candidates selected in the first step were
all visually inspected to make sure that these are real objects and that their
photometry is not corrupted for any complicated reasons that cannot be handled
by SExtractor. During this step, any extremely compact sources would be flagged
as possible brown dwarfs; we found no such source in our case here. Following
Yan et al. (2006), we also used the GOODS {\it Spitzer} IRAC images for further
rejection of possible contaminators from the red galaxy populations at low
redshifts (``IRAC-selected Extremely Red Objects"; Yan et al. 2004). For all
the non-blended cases in the IRAC images where we would have no ambiguity
in identifying counterparts, we found no such contaminators.
A total of 21 candidates have survived after this culling process.

   Fig. 6 shows the image stamps of these $z_{850}$-dropouts, and Table 1 lists
their positions and photometry. These sources have $Y_{105}$ magnitudes
ranging from 26.20 to 29.15 mag. All but one of them have
$z_{850}-Y_{105}\geq 1.0$ mag, indicating that they are likely at
$z\gtrsim 6.6$.  We note that four objects in Table 1 have been previously 
selected as $z_{850}$-dropouts by various groups using the HUDF NIC3 data.
ID zdrop-A032, the second brightest in Table 1, is one of the few earliest
$z_{850}$-dropouts reported, and was first published by YW04b
(ID No.3). We will discuss this source further together with some other objects
in Table 1 later in \S 6.

\begin{figure}[tbp]
\setcounter{figure}{5}
\centering
\caption{ACS and WFC3/IR image stamps of $z\approx 7$ galaxy candidates in 
the HUDF, centered on the source locations. These objects are selected as
$z_{850}$-band dropouts. Images are $2.7^{''}\times 2.7^{''}$ in size; north
is up and east is left. The red circles are 0.5$^{"}$ in radius. Object A032
shows clear double-nucleus morphology, which is indicative of a merging event.
}
\end{figure}

\begin{figure}
\setcounter{figure}{5}
\centering
\includegraphics[width=9.0cm]{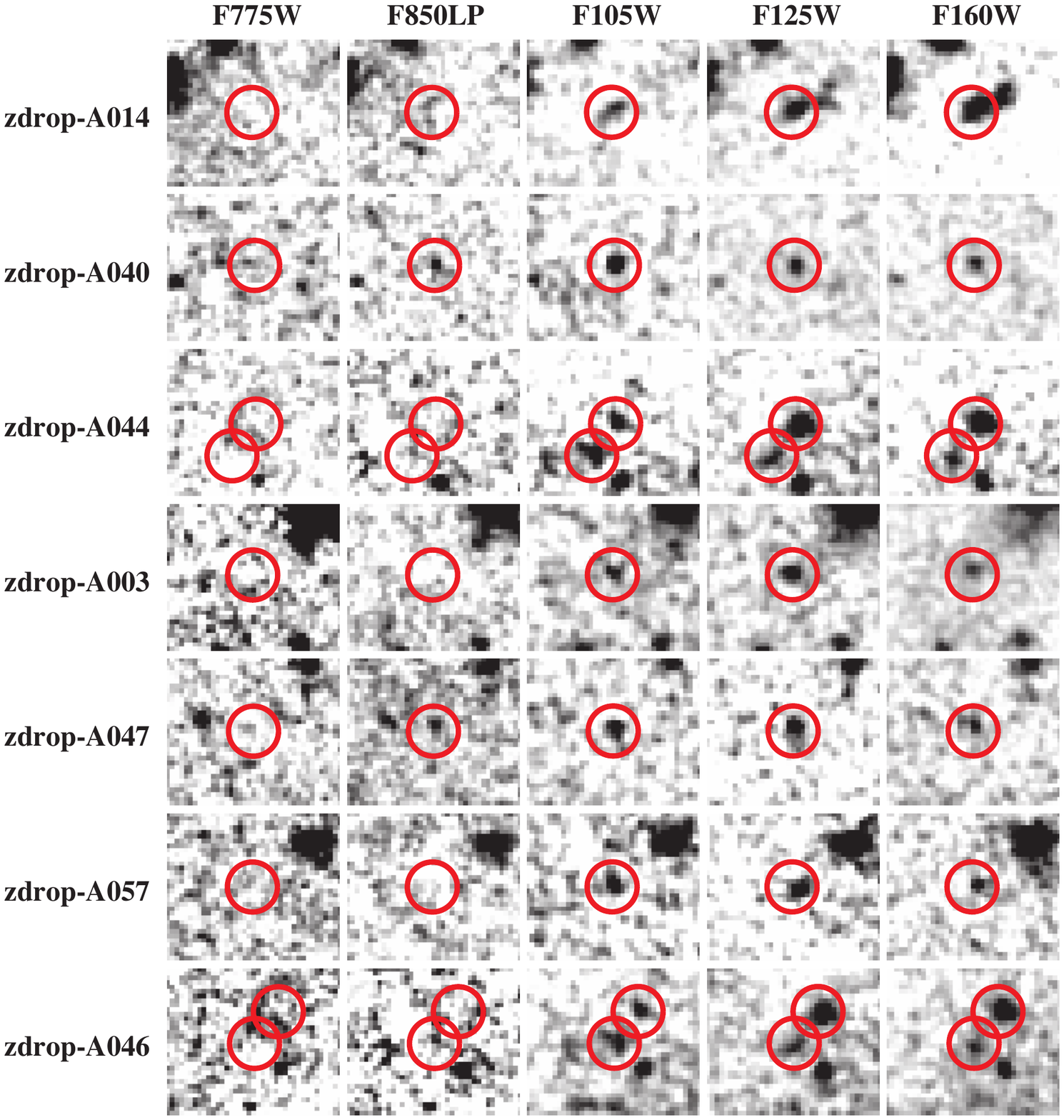}
\caption{(cont.) Note that ID A044 and 046 are close neighbors, which might be
physically associated.}
\end{figure}

\begin{figure}
\setcounter{figure}{5}
\centering
\includegraphics[width=9.0cm]{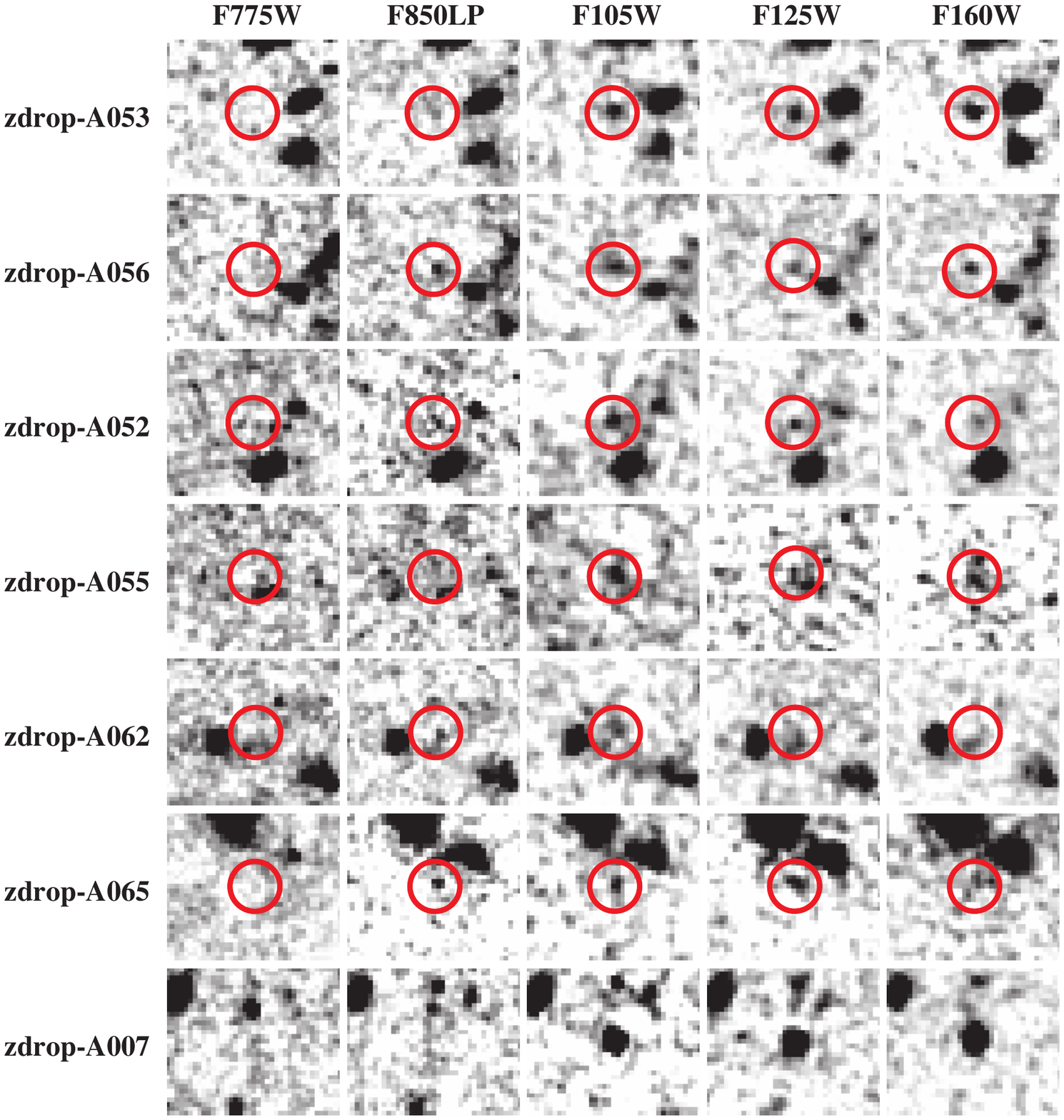}
\caption{(cont.) Note that ID zdrop-A007 probably
is a transient object rather than a genuine $z\approx 7$ galaxy; it is shown
here for completeness.}
\end{figure}

\begin{deluxetable}{lcccccrrrl}
\rotate
\tablewidth{0pt}
\tablecolumns{10}
\tabletypesize{\scriptsize}
\tablecaption{Properties of Galaxy Candidates at $Z\approx 7$\tablenotemark{a}}
\tablehead{
\colhead{ID} &
\colhead{RA \& DEC(J2000)} &
\colhead{$z_{850}$} &
\colhead{$Y_{105}$} &
\colhead{$J_{125}$} &
\colhead{$H_{160}$} &
\colhead{z$-$Y} &
\colhead{Y$-$J} &
\colhead{J$-$H} &
\colhead{Cross.Ref. \& ID\tablenotemark{b}}
}
\startdata

 zdrop-A032\tablenotemark{c} & 3:32:42.560 -27:46:56.593 & 27.94$\pm$0.17 &  26.27$\pm$0.06 & 26.12$\pm$0.04 & 26.14$\pm$0.04 &    1.67  &  0.15 & -0.02 & 1(42566566); 3(zD1); 4(688) \\
 zdrop-A025\tablenotemark{d} & 3:32:38.798 -27:47:07.120 &   $ > 29.95 $  &  27.19$\pm$0.21 & 26.86$\pm$0.12 & 26.74$\pm$0.10 & $> 2.76$ &  0.33 &  0.12 & 1(38807073); 3(zD2); 4(835) \\
 zdrop-A008\tablenotemark{d,e} & 3:32:42.563 -27:47:31.578 &   $ > 29.93 $  &  27.32$\pm$0.27 & 26.77$\pm$0.12 & 26.84$\pm$0.12 & $> 2.61$ &  0.55 & -0.07 & 1(42577314); 3(zD3); 4(1144) \\
 zdrop-A060 & 3:32:43.147 -27:46:28.474 &   $ > 29.80 $  &  27.59$\pm$0.20 & 28.04$\pm$0.22 & 27.69$\pm$0.15 & $> 2.21$ & -0.45 &  0.35 & 1(43146285); 3(zD5); 4(1678) \\
 zdrop-A017 & 3:32:41.044 -27:47:15.529 &   $ > 29.96 $  &  27.90$\pm$0.26 & 28.01$\pm$0.22 & 28.20$\pm$0.24 & $> 2.06$ & -0.11 & -0.19 & 1(41057156); 4(2066) \\
 zdrop-A016 & 3:32:36.380 -27:47:16.199 & 29.41$\pm$0.45 &  28.04$\pm$0.21 & 28.40$\pm$0.22 & 28.37$\pm$0.20 &    1.37  & -0.36 &  0.03 & 1(36387163); 3(zD6); 4(1958) \\
 zdrop-A033 & 3:32:39.577 -27:46:56.449 & 29.95$\pm$0.53 &  28.20$\pm$0.18 & 28.36$\pm$0.16 & 29.10$\pm$0.28 &    1.75  & -0.16 & -0.74 & 1(39586565); 4(1915)  \\
 zdrop-A014\tablenotemark{d} & 3:32:39.556 -27:47:17.534 &   $ > 29.87 $  &  28.20$\pm$0.20 & 27.71$\pm$0.10 & 27.70$\pm$0.09 & $> 1.67$ &  0.49 &  0.01 & 1(39557176); 3(zD4); 4(1092) \\
 zdrop-A040 & 3:32:37.442 -27:46:51.226 & 29.44$\pm$0.30 &  28.21$\pm$0.15 & 28.39$\pm$0.14 & 28.57$\pm$0.15 &    1.23  & -0.18 & -0.18 & 1(37446513); 4(1880) \\
 zdrop-A044\tablenotemark{f} & 3:32:44.703 -27:46:44.245 &   $ > 29.88 $  &  28.21$\pm$0.25 & 27.56$\pm$0.10 & 27.59$\pm$0.10 & $> 1.67$ &  0.65 & -0.03 & 1(44716442); 3(zD7); 4(1107) \\
 zdrop-A003 & 3:32:37.209 -27:48:06.106 &   $ > 29.91 $  &  28.61$\pm$0.33 & 27.89$\pm$0.12 & 28.21$\pm$0.16 & $> 1.30$ &  0.72 & -0.32 & 1(37228061); 3(zD9); 4(1574) \\
 zdrop-A047 & 3:32:40.563 -27:46:43.590 & 30.15$\pm$0.50 &  28.76$\pm$0.22 & 28.90$\pm$0.20 & 29.33$\pm$0.27 &    1.39  & -0.14 & -0.43 & 1(40566437); 3(zD8); 4(2206) \\
 zdrop-A057\tablenotemark{g} & 3:32:39.730 -27:46:21.295 &   $ > 29.93 $  &  29.15$\pm$0.30 & 29.44$\pm$0.30 & 29.50$\pm$0.30 & $> 0.78$ & -0.29 & -0.06 & 1(39736214); 3(zD10); 4(2502)) \\
\hline
 zdrop-A046\tablenotemark{f} & 3:32:44.739 -27:46:44.879 &   $ > 29.87 $  &  27.78$\pm$0.26 & 29.05$\pm$0.61 & 28.39$\pm$0.30 & $> 2.09$ & -1.27 &  0.66 & 4(2888) \\
 zdrop-A053 & 3:32:38.360 -27:46:11.849 &   $ > 29.93 $  &  28.73$\pm$0.26 & 28.60$\pm$0.17 & 28.74$\pm$0.18 & $> 1.20$ &  0.13 & -0.14 & 3(zD11) \\
 zdrop-A056 & 3:32:38.650 -27:46:16.381 &   $ > 29.93 $  &  28.21$\pm$0.25 & 29.90$\pm$0.85 & 29.23$\pm$0.42 & $> 1.72$ & -1.69 &  0.67 &  \\
 zdrop-A052 & 3:32:41.824 -27:46:11.204 &   $ > 29.90 $  &  28.53$\pm$0.27 & 29.32$\pm$0.43 & 29.23$\pm$0.36 & $> 1.37$ & -0.79 &  0.09 &  \\
 zdrop-A055 & 3:32:42.656 -27:46:22.616 &   $ > 29.90 $  &  28.62$\pm$0.29 &     ---        & 29.20$\pm$0.33 & $> 1.28$ &  ---  &  ---  &  \\
 zdrop-A062 & 3:32:37.345 -27:46:28.510 &   $ > 29.92 $  &  28.80$\pm$0.32 & 29.60$\pm$0.50 & 29.70$\pm$0.50 & $> 1.12$ & -0.80 & -0.10 &  \\
 zdrop-A065 & 3:32:36.519 -27:46:41.995 & 29.98$\pm$0.45 &  28.99$\pm$0.30 & 29.25$\pm$0.29 & 29.73$\pm$0.41 &    0.99  & -0.26 & -0.48 & 4(2940) \\
\hline
 zdrop-A007\tablenotemark{h} & 3:32:34.530 -27:47:35.938 &   $ > 29.93 $  &  26.20$\pm$0.03 & 26.05$\pm$0.02 & 25.88$\pm$0.02 & $> 3.73$ &  0.15 &  0.17 & 1(34537360); 3(zD0) \\

\enddata

\tablenotetext{a.}{All magnitude limits are 2-$\sigma$ limits measured within a $r=0.2^{''}$ aperture.}
\tablenotetext{b.}{Cross References: 1 -- Oesch et al. (2009); 3 -- Bunker et al. (2009); 
   4 -- McLure et al. (2009); numbers in parentheses are their corresponding IDs.}
\tablenotetext{c.}{Detected in the NIC3 HUDF data; first reported in YW04.}
\tablenotetext{d.}{Detected in the NIC3 HUDF data; first reported in Bouwens et al. (2004).}
\tablenotetext{e.}{This object happens to be very close to a $Y_{105}$-dropout
(see Table 2, ID z8-B041); Oesch et al. (2009) seem to treat them as one object.}
\tablenotetext{f.}{These two objects are separated by $0.79^{"}$; using their $z_{ph}$ of
$\sim$ 7.0 (see Table 5), this separation correspondes to a proper length of $\sim$ 4.2~kpc.}
\tablenotetext{g.}{This source is at the border line of selection. See text for discussion.}
\tablenotetext{h.}{This bright object is likely a variable source. See Fig. 6 for its light curve.}

\end{deluxetable}

   Oesch et al. (2009) reported 17 $z_{850}$-dropouts using similar (but
not the same) color criteria, 14 of which are also in our sample. These common
objects are listed in the first part of Table 1. Three of
their dropouts escape our selection for various reasons. One of these objects,
their ID UDFz-38537519, is very faint in all our WFC3 images and falls below our
$S/N\geq 3$ threshold. One other object, their ID UDFz-36777536,
has $z_{850}-Y_{105}=0.73$ mag in our catalog and thus was not selected by
our criteria. The last one, their ID UDFz-37807405, which is likely a
genuine $z_{850}$-dropout, was ``vetoed'' by ACS photometry in
our catalog. This source is very close to the brightest star in the HUDF, and
SExtractor reports $S/N>5$ detections in the ACS $V_{606}$ and $i_{775}$ bands
because the strong light gradient caused by this
star badly skews the background estimate at the source location in the ACS
images, resulting in an error in our ACS photometry for this source.

   Seven of our dropouts are not included in Oesch et al. (2009), all of which
have $Y_{105}<29.0$~mag in our catalog. One of them, zdrop-A053 is included in
the sample of Bunker et al. (2009; their ID zD11). Two other objects, our
zdrop-A046 and A065, are included in the sample of McLure et al. (2009; their
IDs 2888 and 2940, respectively). The remaining four, zdrop-A056, A052, A062,
and A055, are not reported elsewhere. These seven sources are 3.4--4.3~$\sigma$ 
detections in our $Y_{105}$ mosaic, although most of them only have $S/N<3$ in
the current $J_{125}$ and $H_{160}$ data (except zdrop-A065 which is detected at
$S/N=3.7$ in $J_{125}$). We expect that these dropouts will be detected at high
significance once the remaining 68\% of the allocated orbits to program
GO-11563 are taken in 2010.

   McLure et al. (2009) have reported six additional candidates at 
$6.6\leq z\leq 7.8$ that are neither in our $z_{850}$-dropout sample nor in
that of Oesch et al. (2009). One of them is actually in our $J_{125}$-dropout
list, which we will discuss later. Their object ID 2794 is not likely a valid
candidate, as it is clearly seen in our $3\times 3$-rebinned ACS $B_{435}$
image. Two of their sources, IDs 2826 and 2487, did not enter our catalog
because their
locations are at the noisy field edges that were excluded from our photometry.
Their ID 2395 was not selected by us because it has $z_{850}-Y_{105}=0.792$ mag
in our catalog. Lastly, their ID 1064, which is likely a valid candidate,
escaped our selection because its ACS photometry in our catalog was contaminated
by a close neighbor that is not seen in the WFC3 IR image.

\begin{figure}[tbp]
\centering
\includegraphics[width=9.0cm]{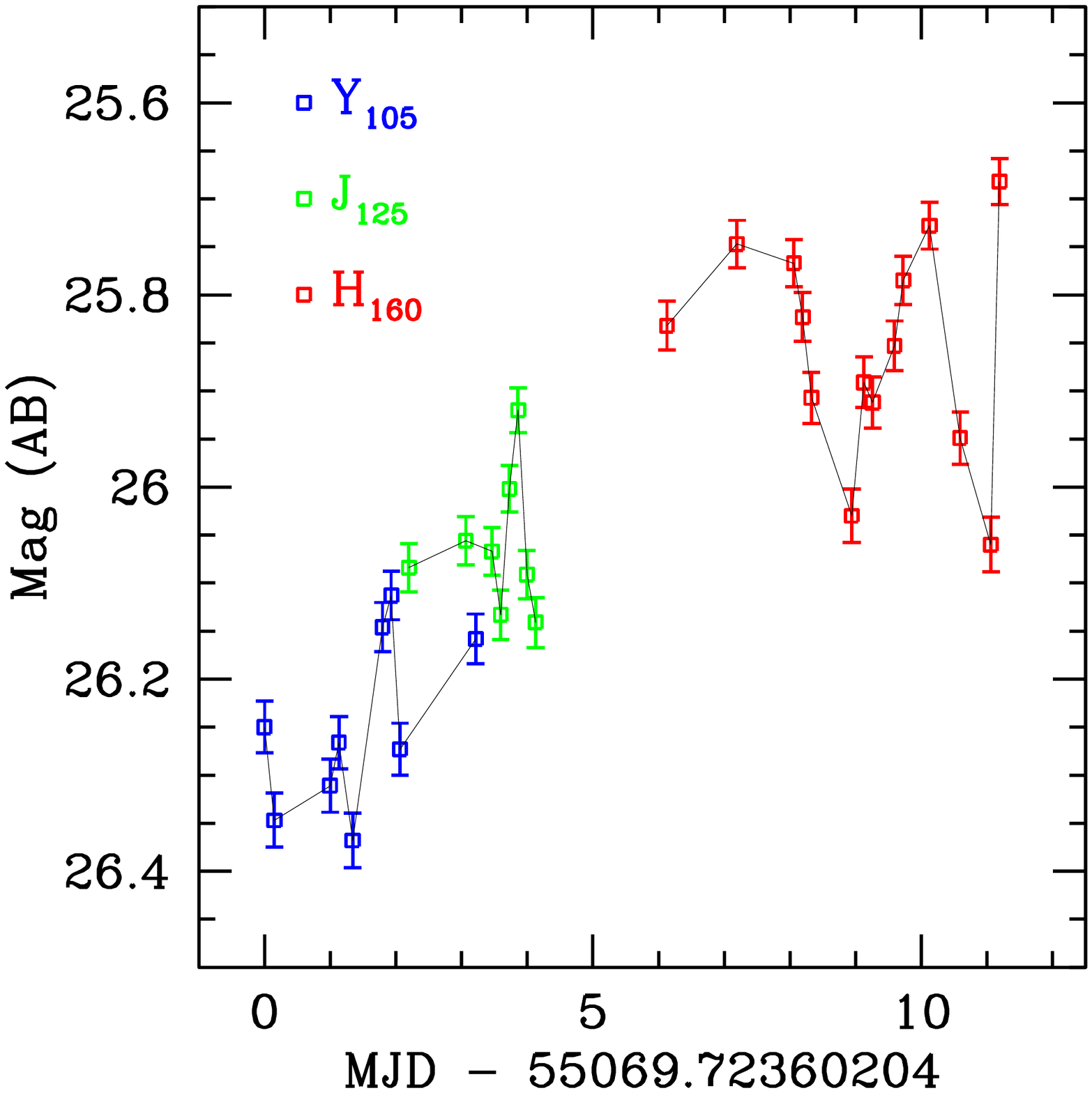}
\caption{The light curves of the possible transient object zdrop-A007. The blue,
green and red data points are its magnitudes in $Y_{105}$, $J_{125}$
and $H_{160}$, respectively. These magnitudes were measured on the 
drizzle-combined mosaics made from the images of individual visits (consisted
of two orbits of exposure each). This object is excluded from further analysis.
}
\end{figure}

   Finally, we comment on the source zdrop-A007, which is the
brightest one in Table 1. As also noted by others, this source is not present
in the HUDF NIC3 images, although its present brightness ($\sim 26$~mag)
is well above the detection limits of those images. Therefore, it must be a
highly variable object. To further explore this possibility, we 
drizzle-combined the images
from each visit separately, and examined this source in each visit. It is
consistent with being a point-source throughout. We measured its magnitudes in
all three passbands as a function of time, using a circular aperture of 5-pixel
in radius. The results are given in Fig. 5, which clearly show that its 
brightness indeed was varying in all three passbands over the period of these
observations. However, the amplitude of variation is $\sim 0.5$~mag at most,
which makes it still difficult to explain its absence from the NIC3 images,
unless it is a transient object. As currently we have no constraint on its
redshift, we exclude this object from further discussion in this paper.

\subsection{Selection of $Y_{105}$-dropouts}

   We first selected $Y_{105}$-dropouts using the $J_{125}$-based catalog.
The main color criterion is $Y_{105}-H_{160}\geq 0.8$ mag, which is appropriate
for identifying the Lyman-break signature at $7.7\leq z\leq 9.4$. 
Valid candidates
should not be present in any ACS passbands at $S/N>2$. No constraint was put on
their $J_{125}-H_{160}$ colors. Objects that meet these criteria were visually
inspected on all the ACS, WFC3 and {\it Spitzer} IRAC images to exclude
contamination from spurious detections and interlopers at low redshifts.
In total, 10 $Y_{105}$-dropouts were selected to $J_{125}\leq 29.0$ mag. All
the five $Y_{105}$-dropouts reported in Bouwens et al. (2009) have been
recovered, and we have found another five more from this analysis.

   As only $\sim 44$\% of the $J_{125}$-band observations have been finished in 
the first epoch, one wonders what deeper observations could add to the result.
Therefore, we also made a composite ``$JH$'' image by weight-combining the
$J_{125}$ and $H_{160}$ mosaics, and used this new ``JH'' image as the
detection image to generate a ``JH''-based catalog for $Y_{105}$-dropout
selection. For an object to be included in this catalog, it was required to
have final extracted $S/N\geq 3$ in either $J_{125}$-band or in $H_{160}$-band.
As it turns out, this $JH$-based catalog includes about 9\% more sources as
compared to the $J_{125}$-based catalog. 

\begin{figure}
\centering
\includegraphics[width=9.0cm]{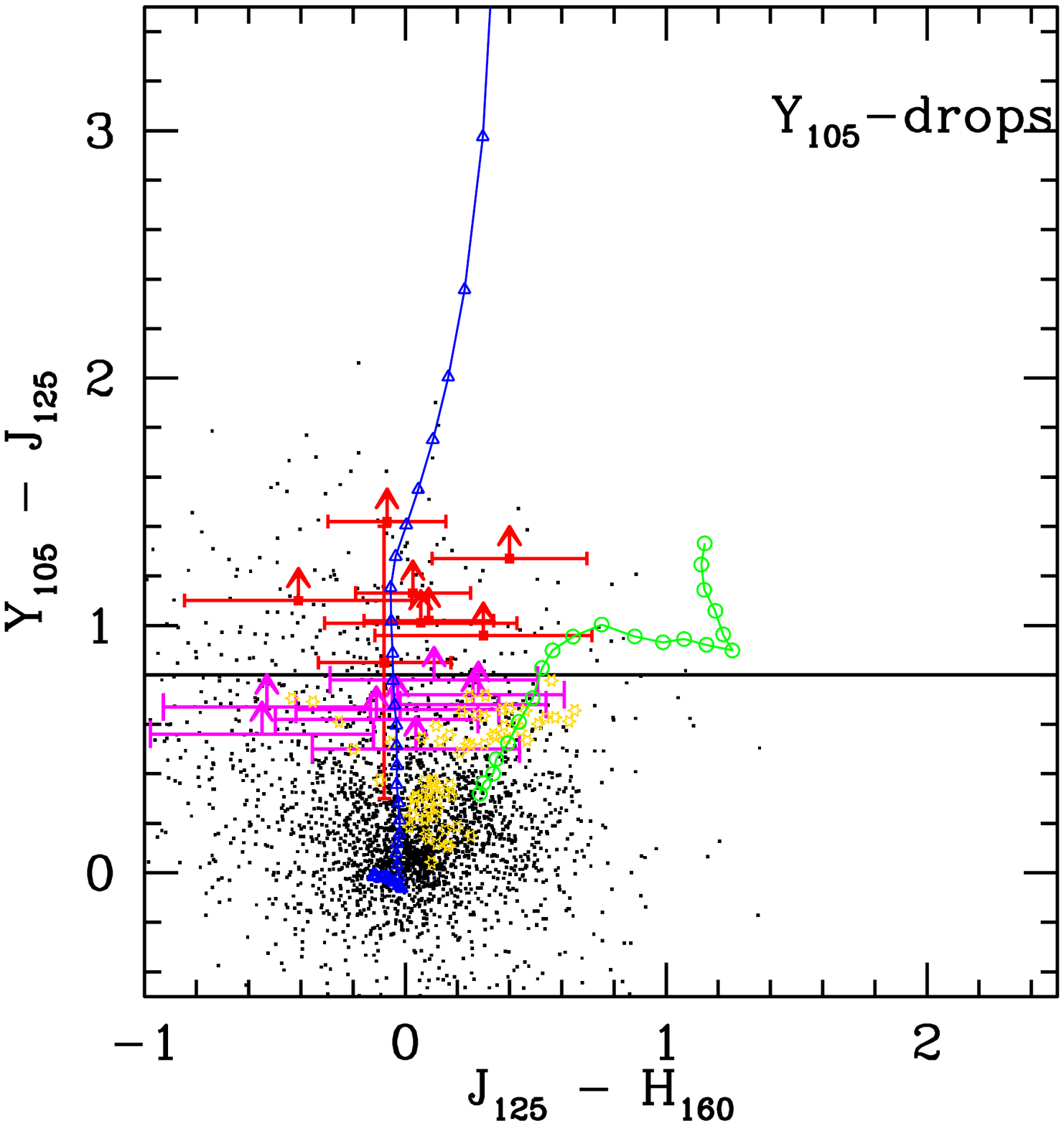}
\caption{Color-color diagram for $Y_{105}$-dropout selection, similar to
Fig. 5. The magenta symbols are the candidates that do not satisfy
$Y_{105}-J_{125}\geq 0.8$~mag when using the 1~$\sigma$ upper limits in
$Y_{105}$ for the non-detections in this band. 
}
\end{figure}

\begin{figure}
\setcounter{figure}{8}
\centering
\includegraphics[width=9.0cm]{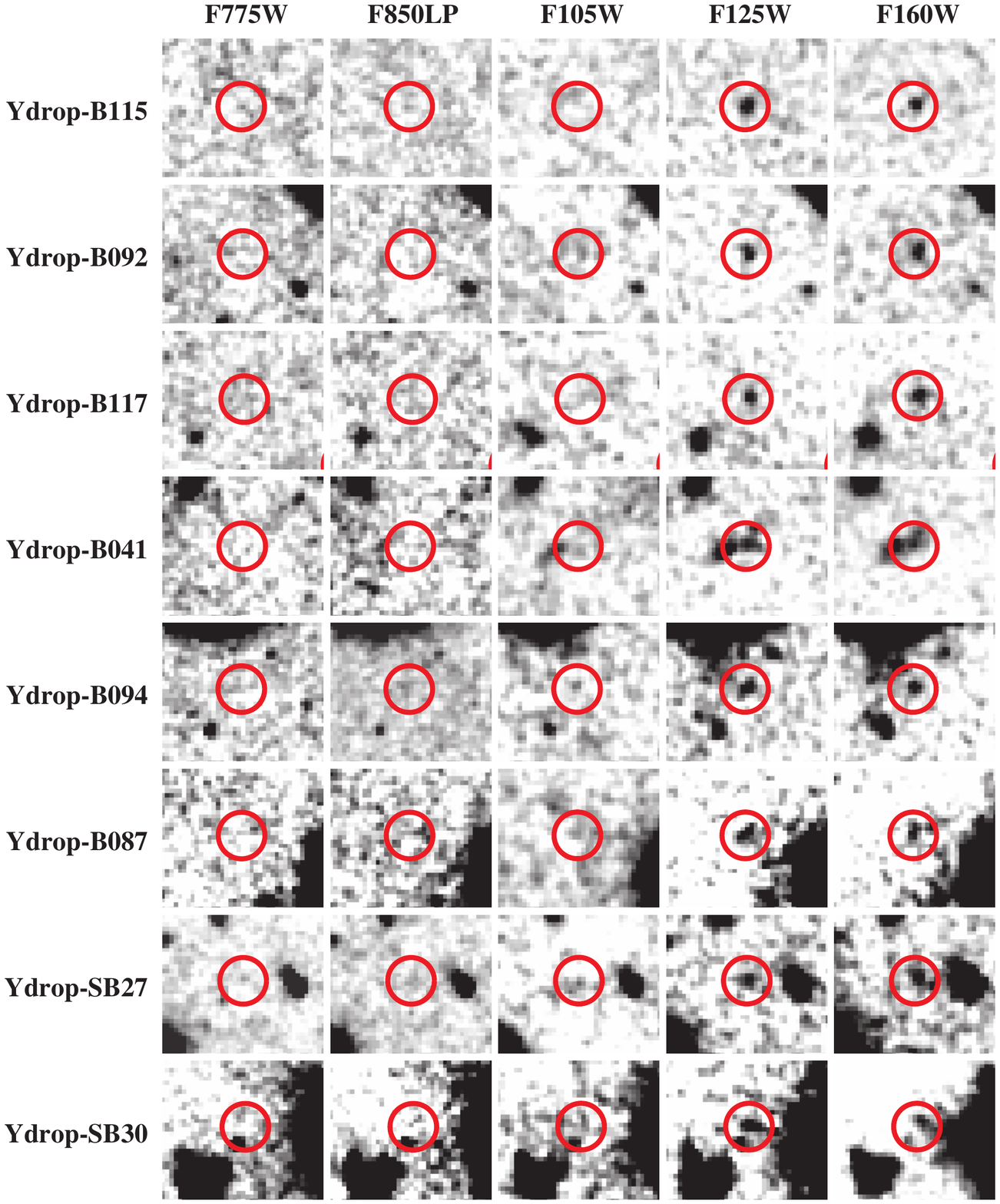}
\caption{Similar to Fig. 6, but for the selection $z\approx 8$ galaxy 
candidates. 
}
\end{figure}

\begin{figure}
\setcounter{figure}{8}
\centering
\includegraphics[width=9.0cm]{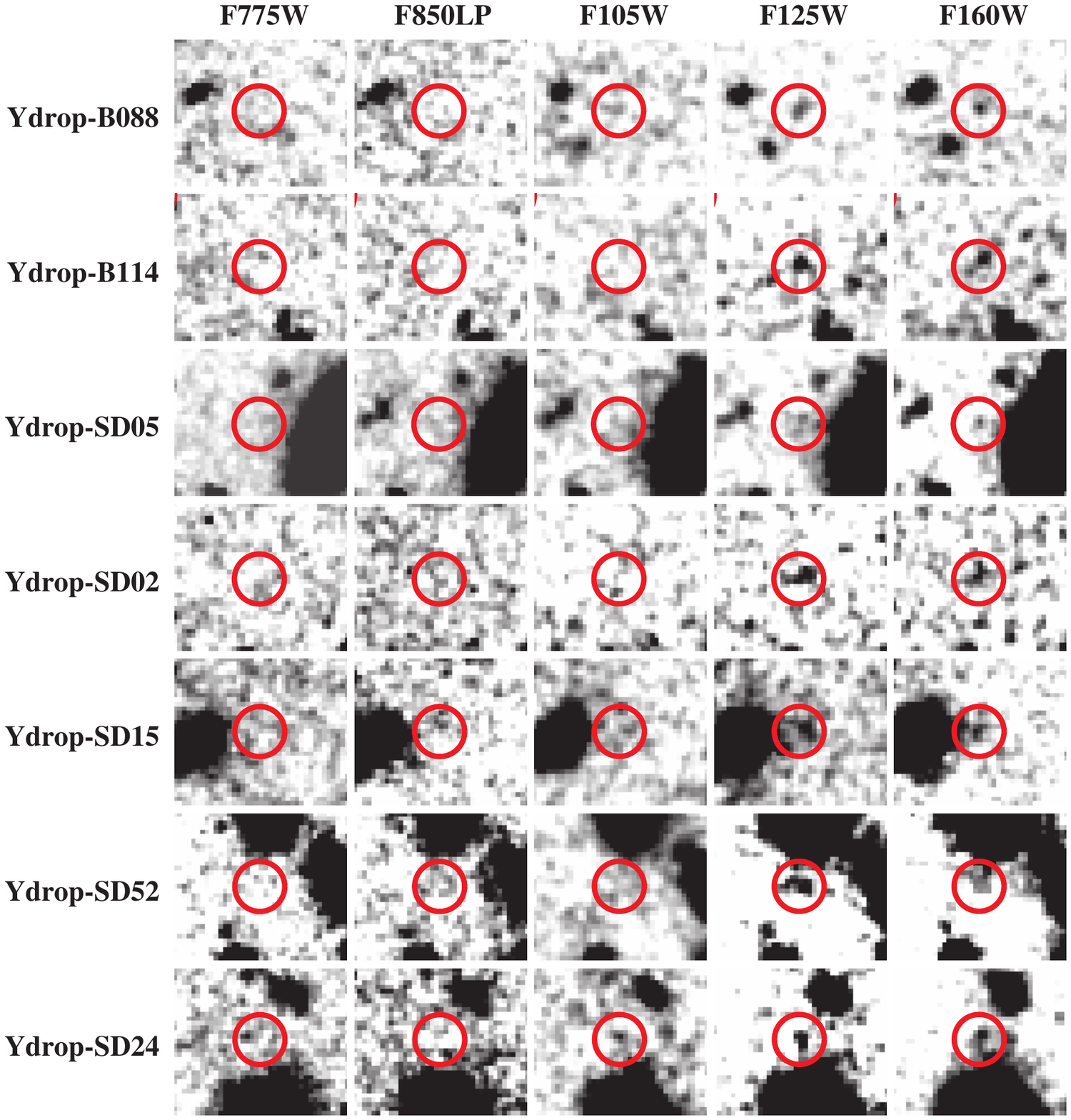}
\caption{(cont.)}
\end{figure}

  We used this $JH$-based catalog to search for $Y_{105}$-dropouts in a similar
way. Somewhat surprisingly, this process resulted in five new candidates. Close
examination shows that they were all in the raw $J_{125}$-based catalog. Two
of them were discarded because they do not have $S/N\geq 3$ as reported
in the $J_{125}$-based catalog. The other three have $S/N>3$ in the
$J_{125}$-based catalog, but their extraction apertures were so defined 
(in the $J_{125}$-band mosaic) that their matched-aperture ACS photometry were
contaminated by unrelated objects nearby and hence erroneously reported
significant detections in the ACS bands. As the result, these three were not
selected in the $J_{125}$-based search when applying our color criteria.
On the deeper $JH$ image, however, the extraction apertures for all these five
objects were better defined, resulted in an improvement in their
photometry that is significant enough to have them selected as candidates.

  Therefore, the total number of $Y_{105}$-dropouts now in our sample is 15. We
point out that, while no constraint to $J_{125}-H_{160}$ color was applied
during selection, all the resulting candidates have 
$J_{125}-H_{160}\lesssim 0.4$~mag.
We show their image stamps in Fig. 6, and list their positions and photometric
properties in Table 2. Their locations in the color space are shown in Fig. 8.
We also note that the formal $Y_{105}-J_{125}$ color limits 
of seven sources (the second part of Table 2, separated by the horizontal line),
calculated using the formal 2-$\sigma$ $Y_{105}$-band flux upper limits within
a $r=0.2^{''}$ aperture, do not strictly meet the $Y_{105}-J_{125}\geq 0.8$~mag
threshold. These include two sources in Bouwens et al., and one different
source in Bunker et al. All the five additional $JH$-based candidates from our
selection fall in this category. However, they only fall slightly short of the
requirement; if we were to use 1-$\sigma$ limit as Bouwens et al. adopted
instead, these sources would certainly go over the threshold. In addition,
their $J_{110}-H_{160}$ colors agree with blue SEDs that are characteristic of
star-forming galaxies. Therefore, we keep them in our sample, but will treat
them differently when interpreting.

   One of these sources, Ydrop-B041, is very close to a $z_{850}$-dropout,
zdrop-A014. Oesch et al. (2009) identify both objects as one single
$z_{850}$-dropout, their UDFz-39557176. Given its color lower limit of
$Y_{105}-J_{125}>1.4$ mag,
we keep it in the $Y_{105}$-dropout sample. McLure et al. (2009) have included
this object (their ID 1422) in their sample as well, and also separate it from
B041 (their ID 1144) as we do. The other seven objects are not reported
elsewhere. 

   Two out of the seven $Y_{105}$-dropouts in Bunker et al. (2009) are neither
in the sample of Bouwens et al. (2009) nor in ours. One object, their
YD04, is too close to the noisy field edge and did not enter our catalog. Their
YD06, did not match with any object in our catalog within a 0.18$^{''}$-radius
(or 2 pixels in our image). The closest object in our catalog lies 0.21$^{''}$
away; this object has a weak detection in the rebinned ACS $B_{435}$ and
$V_{660}$ images and thus did not enter our sample. McLure et al. (2009) also
report an additional source (their ID 2487) at $z\approx 7.8$; however this
source is too close to the noisy field edge that is not considered in our
photometry.

\begin{deluxetable}{lccccccrrrl}
\rotate
\tablewidth{0pt}
\tablecolumns{10}
\tabletypesize{\scriptsize}
\tablecaption{Properties of Galaxy Candidates at $Z\approx 8$\tablenotemark{a}}
\tablehead{
\colhead{ID} &
\colhead{RA \& DEC(J2000)} &
\colhead{$z_{850}$} &
\colhead{$Y_{105}$} &
\colhead{$J_{125}$} &
\colhead{$H_{160}$} &
\colhead{z$-$Y} &
\colhead{Y$-$J} &
\colhead{z$-$J} &
\colhead{J$-$H} &
\colhead{Cross.Ref. \& ID\tablenotemark{b}}
}
\startdata

  Ydrop-B115 & 3:32:38.137  -27:45:54.018 & $ > 29.92 $ &  $ > 29.52 $   &  28.39$\pm$0.17 & 28.36$\pm$0.14 &   ---      & $ > 1.13 $ & $ > 1.53 $ &  0.03 & 2(38135539); 3(YD3); 4(1721) \\
  Ydrop-B092 & 3:32:42.876  -27:46:34.525 & $ > 29.79 $ & 29.30$\pm$0.52 &  28.45$\pm$0.18 & 28.53$\pm$0.18 & $ > 0.49 $ &     0.85   & $ > 1.34 $ & -0.08 & 2(42886345); 3(YD1); 4(1765) \\
  Ydrop-B117 & 3:32:37.800  -27:46:00.149 & $ > 29.92 $ &  $ > 29.51 $   &  28.49$\pm$0.19 & 28.40$\pm$0.16 &   ---      & $ > 1.02 $ & $ > 1.43 $ &  0.09 & 2(37796000); 3(YD2); 4(1939) \\
  Ydrop-B041\tablenotemark{c} & 3:32:39.514  -27:47:17.387 & $ > 29.84 $ &  $ > 29.50 $   &  28.08$\pm$0.16 & 28.15$\pm$0.16 &   ---      & $ > 1.42 $ & $ > 1.76 $ & -0.07 & \\
  Ydrop-B094 & 3:32:43.407  -27:46:36.077 & $ > 29.95 $ &  $ > 29.49 $   &  28.22$\pm$0.25 & 27.82$\pm$0.16 &   ---      & $ > 1.27 $ & $ > 1.73 $ &  0.40 & \\
  Ydrop-B087 & 3:32:42.416  -27:46:24.301 & $ > 29.93 $ &  $ > 29.51 $   &  28.41$\pm$0.26 & 28.82$\pm$0.35 &   ---      & $ > 1.10 $ & $ > 1.52 $ & -0.41 & \\
  Ydrop-SB27 & 3:32:35.172  -27:47:16.966 & $ > 29.82 $ &  $ > 29.55 $   &  28.54$\pm$0.28 & 28.48$\pm$0.24 &   ---      & $ > 1.01 $ & $ > 1.28 $ &  0.06 & \\
  Ydrop-SB30 & 3:32:34.307  -27:47:11.476 & $ > 29.85 $ &  $ > 29.51 $   &  28.55$\pm$0.34 & 28.25$\pm$0.24 &   ---      & $ > 0.96 $ & $ > 1.30 $ &  0.30 & \\
\hline
  Ydrop-B088 & 3:32:43.081  -27:46:27.714 & $ > 29.88 $ &  $ > 29.53 $   &  28.86$\pm$0.22 & 29.39$\pm$0.33 &   ---      & $ > 0.67 $ & $ > 1.02 $ & -0.53 & 2(43086276); 4(2841) \\
  Ydrop-B114 & 3:32:37.635  -27:46:01.571 & $ > 29.93 $ &  $ > 29.50 $   &  28.94$\pm$0.23 & 29.49$\pm$0.36 &   ---      & $ > 0.56 $ & $ > 0.99 $ & -0.55 & 2(37636015); 3(YD7); 4(2079) \\
  Ydrop-SD05 & 3:32:36.532  -27:47:50.258 & $ > 29.82 $ &  $ > 29.52 $   &  28.74$\pm$0.31 & 28.63$\pm$0.25 &   ---      & $ > 0.78 $ & $ > 1.08 $ &  0.11 & \\
  Ydrop-SD02 & 3:32:37.583  -27:48:00.245 & $ > 29.95 $ &  $ > 29.48 $   &  28.80$\pm$0.23 & 28.54$\pm$0.16 &   ---      & $ > 0.68 $ & $ > 1.15 $ &  0.26 & \\
  Ydrop-SD15 & 3:32:38.409  -27:47:25.098 & $ > 29.95 $ &  $ > 29.56 $   &  28.84$\pm$0.27 & 28.56$\pm$0.19 &   ---      & $ > 0.72 $ & $ > 1.11 $ &  0.28 & 3(YD5) \\
  Ydrop-SD52 & 3:32:38.809  -27:47:16.188 & $ > 29.90 $ &  $ > 29.52 $   &  28.90$\pm$0.28 & 29.01$\pm$0.27 &   ---      & $ > 0.62 $ & $ > 1.00 $ & -0.11 & \\
  Ydrop-SD24 & 3:32:35.845  -27:47:17.156 & $ > 29.95 $ &  $ > 29.54 $   &  29.04$\pm$0.30 & 29.00$\pm$0.26 &   ---      & $ > 0.50 $ & $ > 0.91 $ &  0.04 & \\

\enddata

\tablenotetext{a.}{All magnitude limits are 2-$\sigma$ limits measured within a $r=0.2^{''}$ aperture.}
\tablenotetext{b.}{Cross References: 2 -- Bouwens et al. (2009); 3 -- Bunker et al. (2009); 4 -- McLure et al. (2009);
 numbers in parentheses are their corresponding IDs.}
\tablenotetext{c.}{This object is consisted of two components. It is very close to our ID z7-A008,
   and Oesch et al. (2009) take them as one object (their ID 39557176).} 

\end{deluxetable}

\subsection{Selection of $J_{125}$-dropouts}

   As Fig. 1 illustrates, galaxies at $z\approx 10$ can in principle be
selected as $J_{125}$-dropouts. We carried out such a selection using the
$H_{160}$-based catalog. The main criteria applied are
$J_{125}-H_{160}>0.8$ mag (Fig. 8), and $S/N<2$ in the $Y_{105}$-band and all
the four ACS bands. This process selects candidates at 
$9.4\lesssim z\lesssim 11.8$. As before, we visually inspected the candidates
on all the ACS, WFC3, and IRAC images. 
In total, our selection has resulted in 20 $J_{125}$-dropouts to
$H_{160}=29.0$ mag; all have $S/N=3.0$--4.9 in the $H_{160}$-band. We list their
locations and photometry in Table 3, and show their image stamps in Fig. 7.

\begin{figure}[tbp]
\centering
\includegraphics[width=9.0cm]{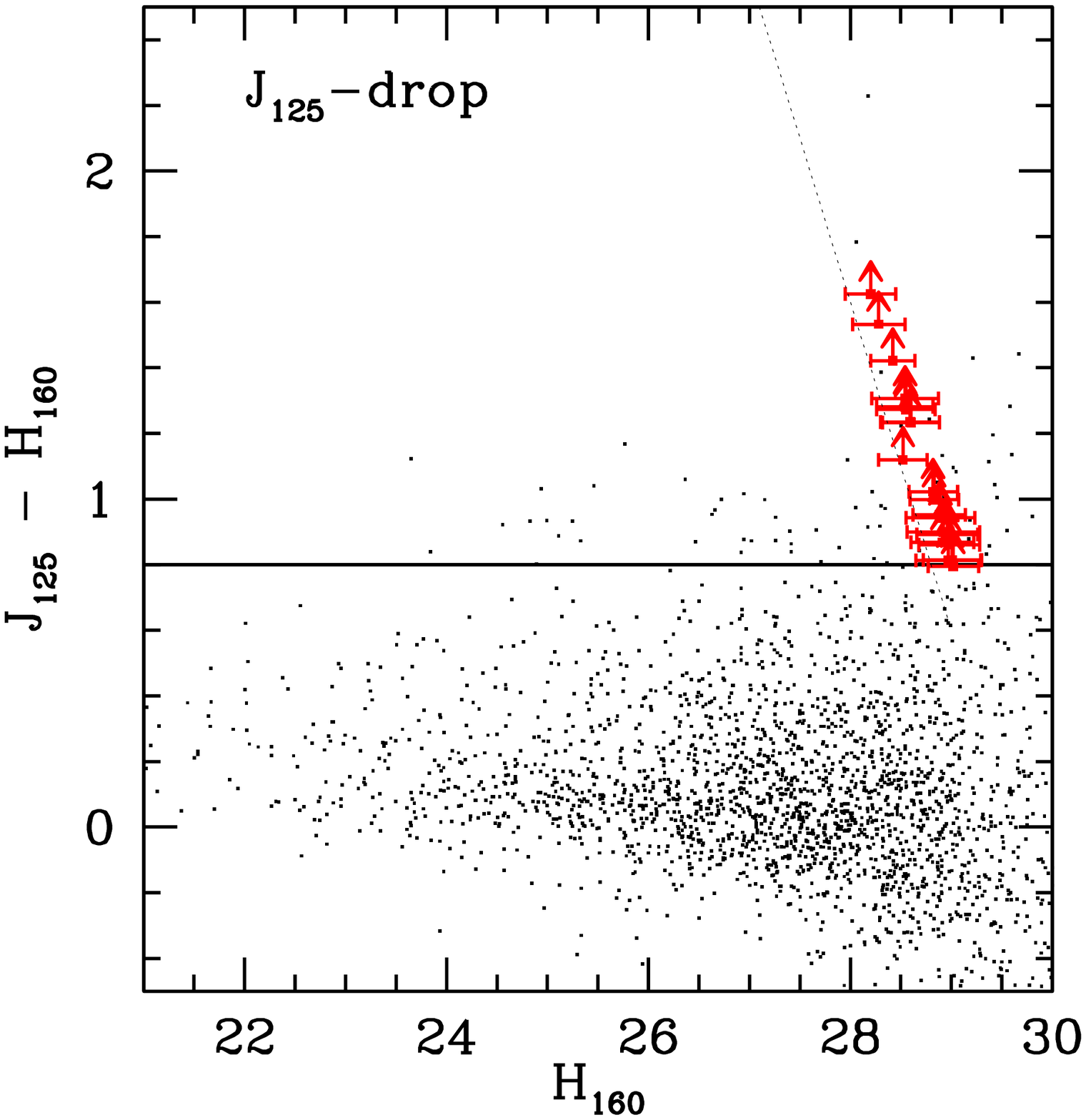}
\caption{Color-magnitude diagram for $J_{125}$-dropout selection. The
black dots are field objects, and the horizontal line indicates the major
color criteria of $J_{125}-H_{160}\geq 0.8$~mag. The red data points are
the selected candidates. Their positions are tilted (as indicated by the
thin dashed line) because their
2-$\sigma$ limits in $J_{125}$ is $\sim 29.8$~mag.
}
\end{figure}

\begin{figure}
\setcounter{figure}{10}
\centering
\includegraphics[width=9.0cm]{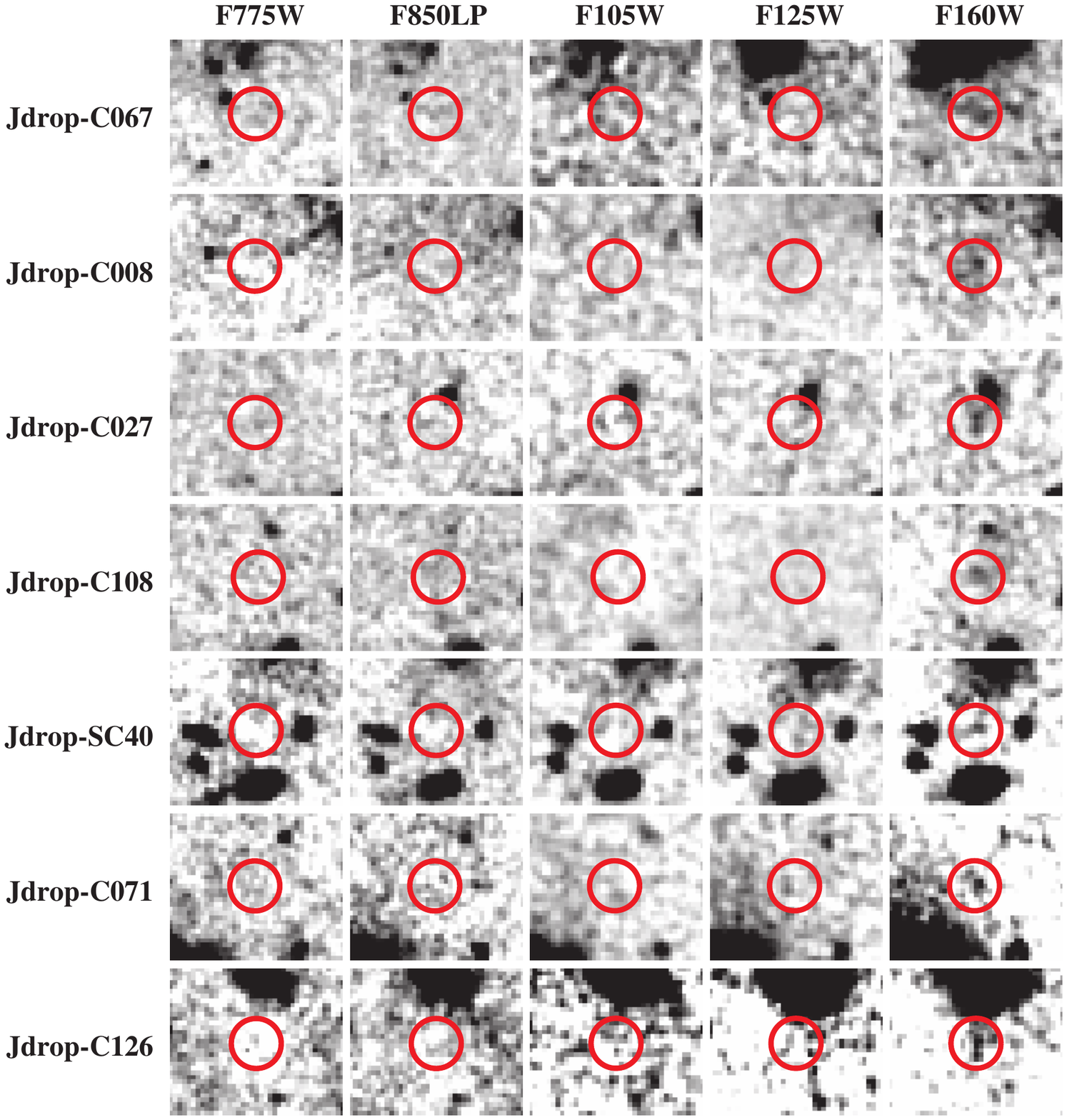}
\caption{Similar to Fig. 3 \& 4, but for $z\approx 10$ galaxy candidates. 
Objects are selected as $J_{125}$-dropouts.
}
\end{figure}

\begin{figure}
\setcounter{figure}{10}
\centering
\includegraphics[width=9.0cm]{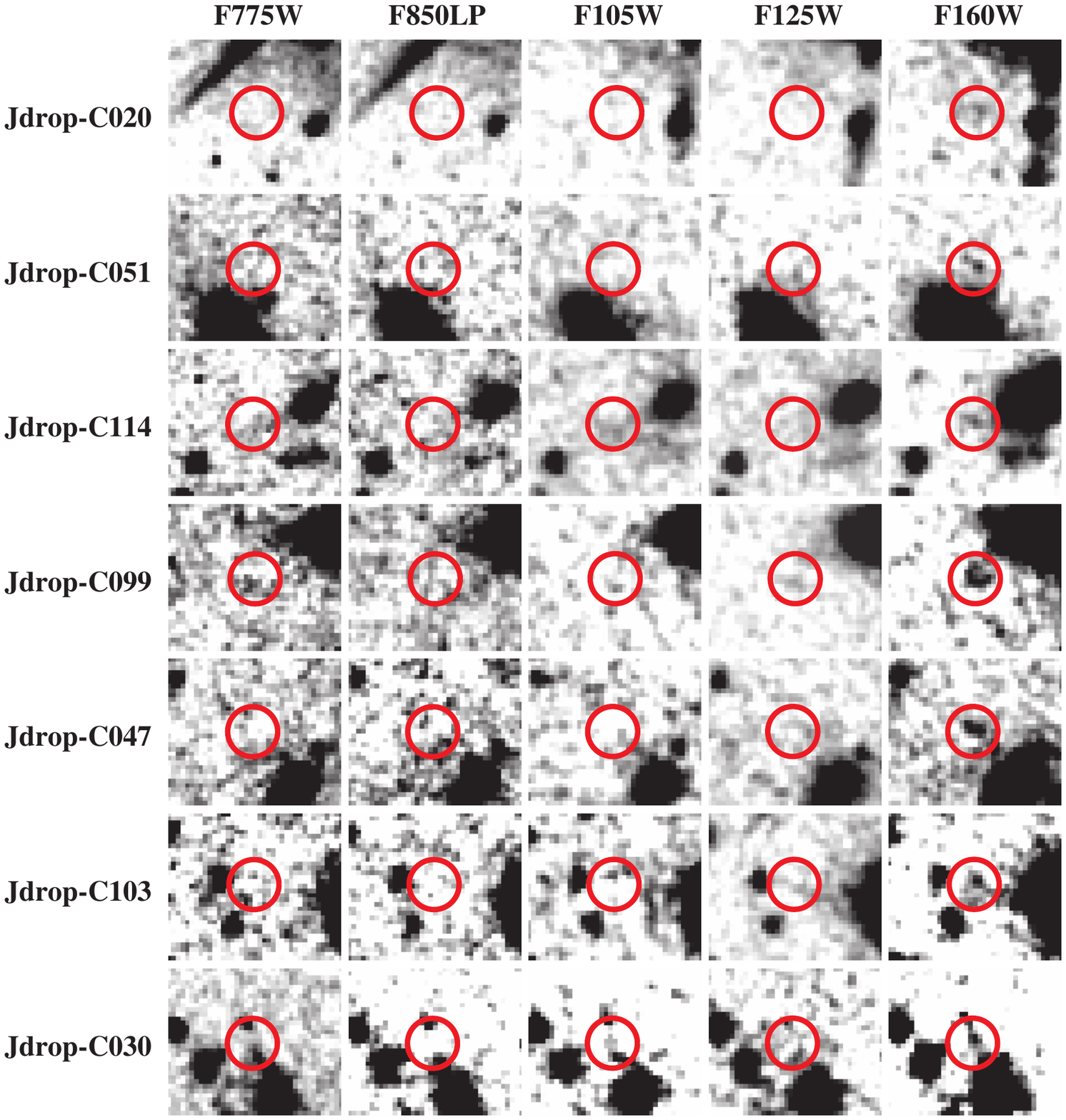}
\caption{(cont.)
}
\end{figure}

\begin{figure}
\setcounter{figure}{10}
\centering
\includegraphics[width=9.0cm]{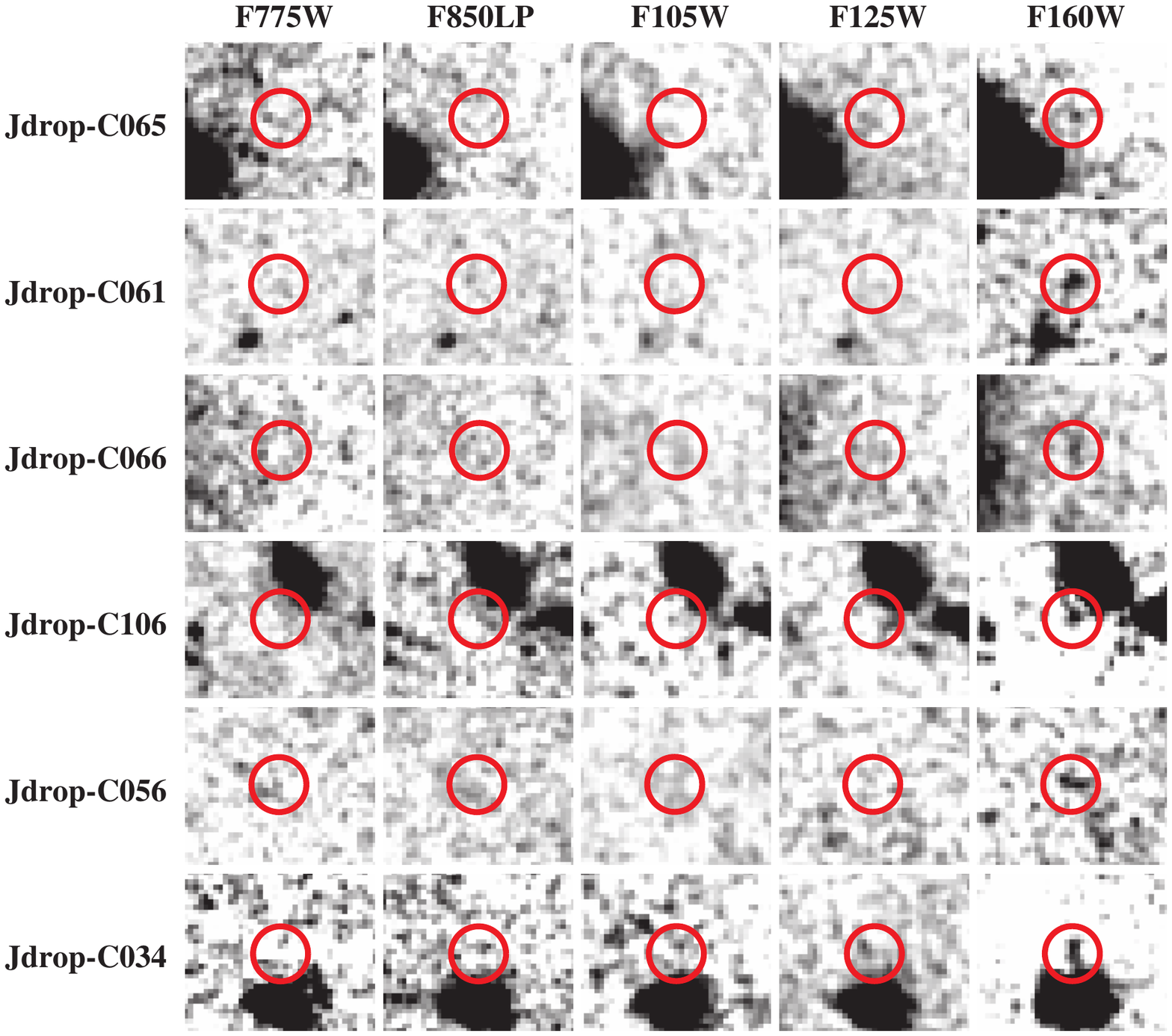}
\caption{(cont.)
}
\end{figure}

   While we only required that $J_{125}-H_{160}>0.8$ mag, all these objects are 
undetected in the current $J_{125}$ mosaic, making them single-band detections. 
One major concern about any single-band detections at or approaching the
detection limit is whether they are real objects at all. In order to address
this question, we performed two tests. We first visually examined every
individual, distortion-corrected $H_{160}$-band image and its associated pixel
mask at the location of each candidate to see whether any of these objects
could originate from residual cosmic-ray hits, other image defects, or any 
transients (such as asteroids). We did not find any evidence that this could be
the case. We then broke the image set into two groups, each consisting of 28
images, and ran MultiDrizzle on these two subsets separately. This was done
twice, first splitting the images by visits (i.e., the first group was made of
the images from the first seven visits and the second group was made of the 
images from the latter seven visits) and then splitting them by taking out half
of the images from each visit. We then inspected the locations of our 
candidates, and found that they could still be seen in these mosaics that
contain only half of the available data, albeit being weaker.

\begin{figure}
\centering
\includegraphics[width=9.0cm]{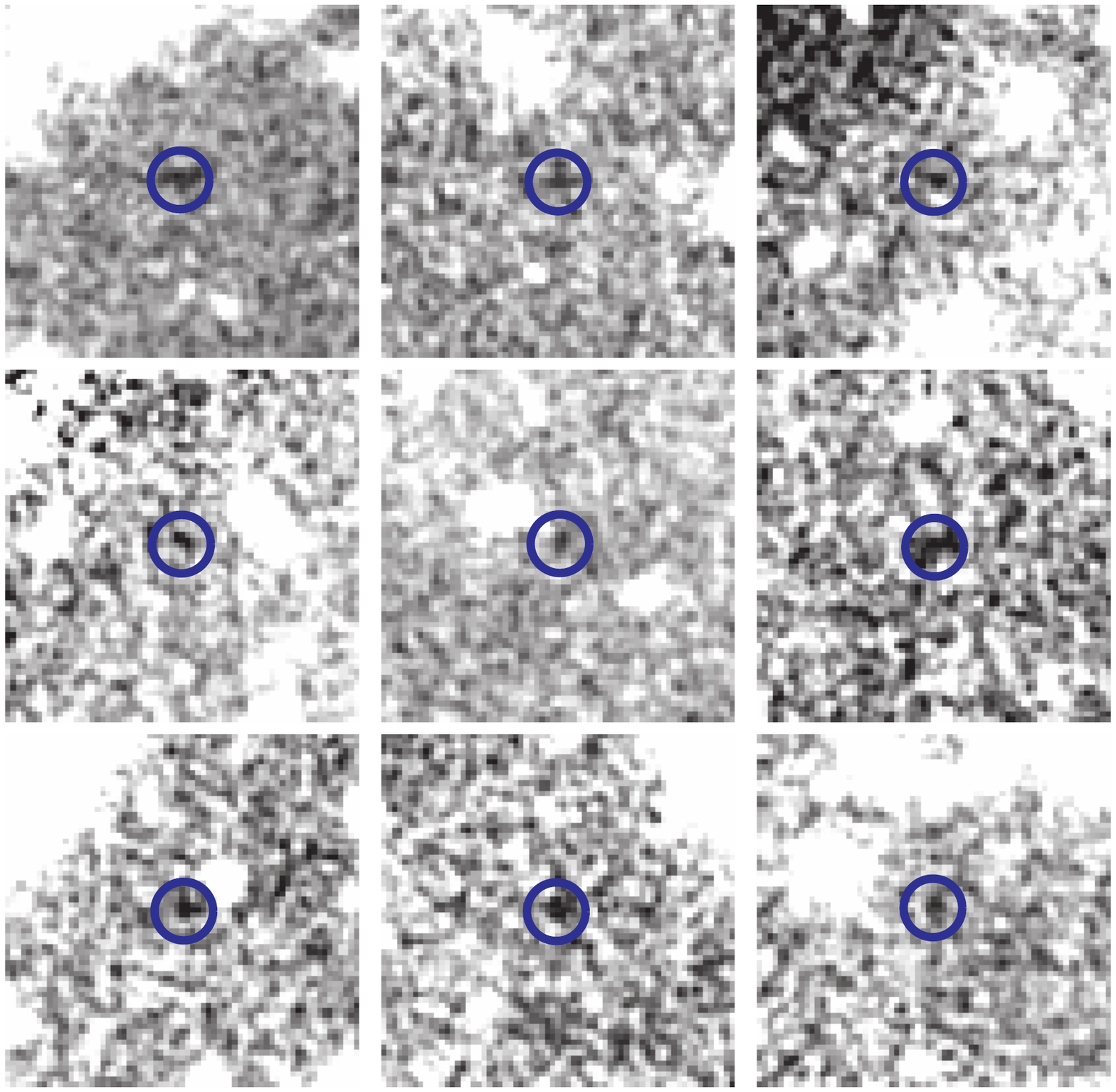}
\caption{Negative $H_{160}$-band image cutouts of the nine ``negative
candidates" that pass our $J_{125}$-dropout selection criterion in the 
``negative image" test. The cutouts are $6^{''}\times 6^{''}$ in size. The blue
circles are 0.5$^{''}$ in radius. The test suggests that our $J_{125}$-dropout
sample could be contaminated at the level of 33--47\% due to noise fluctuation.
}
\end{figure}

   These tests strongly suggest that the $H_{160}$-band detections of our
$J_{125}$-dropouts are the cumulative results of the weak but positive counts in
the individual $H_{160}$-band images, which is characteristic of real objects.
Strictly speaking, however, these tests still do not answer whether such a
detection could be the cumulative result of the noise fluctuations in 
individual images that happen to be mostly positive in a few connected pixels.
To address this issue,
we carried out the so-called ``negative image" test (e.g., Dickinson et al.
2004). The $H_{160}$-band mosaic, which has the mean background of zero by the
way it was produced (see \S 3.1), was multiplied by $-1$ to make any positive
pixels to negative and vise versa. We then extracted
``sources" in this negative image by running SExtractor in the same way as we
did with the real $H_{160}$ mosaic, including using the same RMS map.
As the noise fluctuation has the same probability of scattering a pixel to
a negative value below the mean or to a posistive value above the mean, the
occurance of the ``sources" in the negative image would represent that of the
spurous detections due to the noise fluctuation in the original image. The
detected negative ``sources" were further examined as they were real candidate
dropouts, and we found that nine of them would meet our selection criteria.
The image cutouts of these nine ``negative candidates" are shown in Fig. 12.
One of them would have $28.0< H_{160}\leq 28.5$~mag, while the rest would have
$28.5<H_{160}\leq 29.0$~mag. As our $J_{125}$-dropout sample has three objects
in the former magnitude bin and 17 in the latter (see Table 3), these
negative ``dropouts" imply the contamination rates of 33.3\% and 47.1\% in our
sample in these two bins, respectively. We note that these are significantly
lower than Bouwens et al. (2010) estimated using a similar sets of extraction 
parameters on their mosaics.

\begin{deluxetable}{ccccccc}
\tablewidth{0pt}
\tablecolumns{10}
\tabletypesize{\scriptsize}
\tablecaption{Properties of Galaxy Candidates at $Z\approx 10$ \tablenotemark{a}}
\tablehead{
\colhead{ID} &
\colhead{RA \& DEC(J2000)} &
\colhead{$z_{850}$} &
\colhead{$Y_{105}$} &
\colhead{$J_{125}$} &
\colhead{$H_{160}$} &
\colhead{J$-$H}
}
\startdata

  Jdrop-C067 & 3:32:43.291 -27:46:48.022 & $> 29.903 $ & $> 29.523 $ & $> 29.825 $ &  28.20$\pm$0.25 & $> 1.62 $ \\
  Jdrop-C008 & 3:32:39.252 -27:48:02.603 & $> 29.949 $ & $> 29.538 $ & $> 29.813 $ &  28.28$\pm$0.26 & $> 1.53 $ \\
  Jdrop-C027 & 3:32:35.073 -27:47:40.690 & $> 29.944 $ & $> 29.568 $ & $> 29.841 $ &  28.42$\pm$0.22 & $> 1.42 $ \\
  Jdrop-C108 & 3:32:38.870 -27:45:38.632 & $> 29.920 $ & $> 29.267 $ & $> 29.640 $ &  28.52$\pm$0.24 & $> 1.12 $ \\
  Jdrop-SC40 & 3:32:37.450 -27:46:14.250 & $> 29.906 $ & $> 29.486 $ & $> 29.846 $ &  28.54$\pm$0.33 & $> 1.31 $ \\
  Jdrop-C071 & 3:32:39.588 -27:46:09.124 & $> 29.892 $ & $> 29.538 $ & $> 29.821 $ &  28.54$\pm$0.28 & $> 1.28 $ \\
  Jdrop-C126 & 3:32:38.277 -27:46:05.315 & $> 29.930 $ & $> 29.529 $ & $> 29.823 $ &  28.55$\pm$0.29 & $> 1.27 $ \\
  Jdrop-C020 & 3:32:38.127 -27:47:44.819 & $> 29.931 $ & $> 29.451 $ & $> 29.825 $ &  28.59$\pm$0.29 & $> 1.23 $ \\
  Jdrop-C051 & 3:32:37.738 -27:47:04.888 & $> 29.878 $ & $> 29.507 $ & $> 29.833 $ &  28.60$\pm$0.28 & $> 1.23 $ \\
  Jdrop-C114 & 3:32:35.168 -27:46:47.982 & $> 29.926 $ & $> 29.477 $ & $> 29.843 $ &  28.82$\pm$0.24 & $> 1.02 $ \\
  Jdrop-C099 & 3:32:43.614 -27:46:35.162 & $> 29.854 $ & $> 29.517 $ & $> 29.828 $ &  28.83$\pm$0.24 & $> 1.00 $ \\
  Jdrop-C047 & 3:32:41.754 -27:47:12.484 & $> 29.926 $ & $> 29.471 $ & $> 29.832 $ &  28.88$\pm$0.26 & $> 0.95 $ \\
  Jdrop-C103 & 3:32:35.017 -27:46:40.073 & $> 29.884 $ & $> 29.542 $ & $> 29.833 $ &  28.89$\pm$0.34 & $> 0.94 $ \\
  Jdrop-C030 & 3:32:36.648 -27:47:36.758 & $> 29.928 $ & $> 29.501 $ & $> 29.778 $ &  28.91$\pm$0.31 & $> 0.87 $ \\
  Jdrop-C065 & 3:32:42.841 -27:46:48.954 & $> 29.903 $ & $> 29.493 $ & $> 29.820 $ &  28.92$\pm$0.36 & $> 0.90 $ \\
  Jdrop-C061 & 3:32:43.007 -27:46:53.263 & $> 29.915 $ & $> 29.506 $ & $> 29.852 $ &  28.96$\pm$0.30 & $> 0.89 $ \\
  Jdrop-C066 & 3:32:38.493 -27:46:48.608 & $> 29.937 $ & $> 29.473 $ & $> 29.783 $ &  28.97$\pm$0.32 & $> 0.81 $ \\
  Jdrop-C106 & 3:32:36.582 -27:46:41.365 & $> 29.948 $ & $> 29.556 $ & $> 29.841 $ &  28.98$\pm$0.30 & $> 0.86 $ \\
  Jdrop-C056 & 3:32:37.588 -27:46:58.451 & $> 29.939 $ & $> 29.514 $ & $> 29.824 $ &  29.01$\pm$0.29 & $> 0.81 $ \\
  Jdrop-C034 & 3:32:39.255 -27:47:35.758 & $> 29.841 $ & $> 29.469 $ & $> 29.815 $ &  29.02$\pm$0.25 & $> 0.80 $ \\

\enddata
\tablenotetext{a.}{All magnitude limits are 2-$\sigma$ limits measured within a $r=0.2^{''}$ aperture.}

\end{deluxetable}

\subsection{Compare to the recent result of Bouwens et al.}

    Recently, Bouwens et al. (2010) reported the result of their search of
$J_{125}$-dropouts in this field. They have found a total of three 
$J_{125}$-dropouts, none of which are in our sample. They also claimed that the
majority of our $J_{125}$-dropouts are implausibly too close to ``bright, 
foreground galaxies", and hence questioned any of our $J_{125}$-dropouts being
real.

\begin{figure}
\centering
\includegraphics[width=9.0cm]{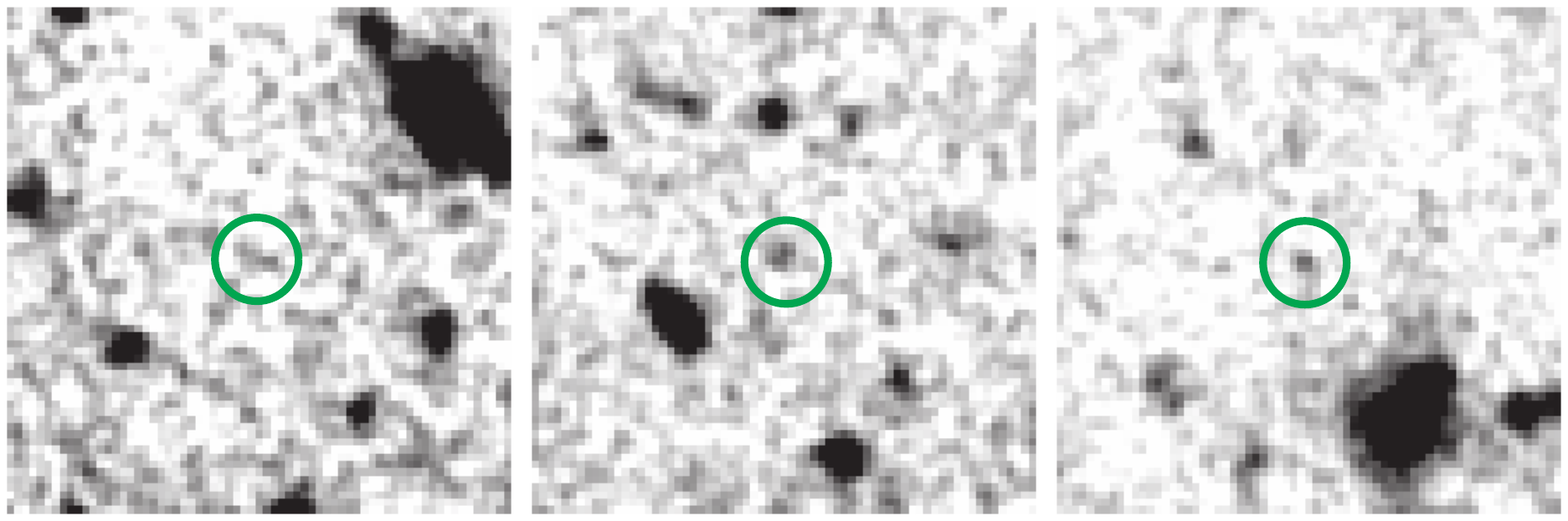}
\caption{Three $J_{125}$-dropouts of Bouwens et al. (2010) as seen in our
$H_{160}$-band image. From left to right, these are their UDFj-43696407,
UDFj-35427336 and UDFj-38116243. The cutouts are $6^{''}\times 6^{''}$ in size;
north is up and east is left. The green circles are 0.5$^{''}$ in radius. The
first object is a non-detection in our image, and the other two have magnitudes
fainter than our selection criteria. See text for details.
}
\end{figure}

    We checked their three $J_{125}$-dropouts in our data. Fig. 13 shows the
cutout images at these locations in our $H_{160}$-band mosaic. Their
UDFj-43696407, which they reported $H_{160}=28.9\pm 0.2$~mag, turns out to be a
formal non-detection in our image, and has the 2-$\sigma$ upper limit within a
$0.2^{''}$-radius aperture of 29.91~mag. The other two objects, UDFj-35427336
and UDFj-38116243, are detected in our image, but have $H_{160}=29.64\pm 0.31$
and $29.70\pm 0.28$~mag, respectively. As these are fainter than our adopted
$J_{125}$-dropout selection threshold of $H_{160}\leq 29.2$~mag (see \S 4.1),
they would not be included in our sample. We note that Bouwens et al. reported
$H_{160}=29.1\pm 0.2$ and $28.9\pm 0.2$~mag for these two objects, respectively.

    We believe that such discrepancies could be partly explained by the
differences in the data reduction, which we already emphasize in \S 3. The two
groups used different softwares and have adopted different approaches to
process the data (e.g., reference files, registration, background subtraction,
etc.), and it is very likely that there are also differences in the adopted
parameters that could affect the results (e.g., drizzle scale, pixel scale,
cosmic-ray rejection, weighting, etc.). It is not unconceivable that some
differences in the mosaics would exist at the faintest level. For this reason,
while the non-detection of one of their candidates is still puzzling, we do not
exclude the possibility that the other two could be $J_{125}$-dropouts. The
exclusion of these two objects from our sample, however, is demanded by our
data and photometry. Even if future data show that their brightness agrees
better with the photometry of Bouwens et al. than to ours, their exclusion from
our current sample still does not affect our major conclusion in \S 6. We know
that the current data are significantly incomplete at $H_{160}\approx 29$~mag
(see Fig. 3), and we expect our sample to be incomplete at this level. And this
is precisely the reason that we have applied the incompleteness correction, as
detailed in \S 6.3.

   Bouwens et al.'s criticism that our candidates are spurious, however, is
flawed. Their Fig. 7, based on which they claimed that our candidates are
predominantly close to bright foreground sources, is not justified. It is
not clear to us how they define ``bright foreground sources", ``blank sky",
or ``distance". While their ``relative likelihood" is also not clearly defined,
we believe that it is similar to histogram normalized by total counts (but not
entirely the same, as their ``relative likelihood" does not add up to unity).
Regardless of these ambiguities, their Fig. 7 concludes that the majority of our
$J_{125}$-dropouts lie within $0.5^{''}$ to ``bright, foreground galaxies". We
disagree with their assessment. In the image montages of the candidates
shown in our current paper, including Fig. 11 for the $J_{125}$-dropouts, the
red circles used to indicate the source locations are always $0.5^{''}$
in radius, and one can see that the majority of our candidates do not have
their red circles overlapped with any neighbor. To show this more clearly,
Fig. 14 display the $H_{160}$ image cutouts of all the 20 $J_{125}$-dropouts,
in the order of their appearance in Table 3 (i.e., in decreasing $H_{160}$-band
brightness). The cutouts are $6^{''}\times 6^{''}$ in size, which is large
enough to see the neighbors. We count six objects (30\%) that have their red
circles making contact with a bright neighbour (C106, C034, C030, SC40, C114,
and C126), while Bouwens et al.'s Fig. 7 indicates that $\sim$~90\% of our
objects are within $0.5^{''}$ to a bright neighbor, which is misleading.
Similarly, their same criticism to our $Y_{105}$-dropouts that are not in
common with theirs is flawed as well. We count only three objects in our
$Y_{105}$-dropout sample that have their red circles making contact with a
bright neighbour (SD05, SD15 and SD52). In fact, one of these three objects,
SD05, is also selected by Bunker et al. (2009) as a candidate (see Table 2). 

    To better address whether our $J_{125}$-dropouts are preferentially close
to foreground neighbors, we performed the following test. We generated 6,239
random positions in the field, and computed the distances between these
positions to the centroids of the nearest objects in the $H_{160}$-based
catalogs. Similarily, we also calculated the distances between the
$J_{125}$-dropouts to the centroids of their nearest neighbors. The
distributions of these distances are normalized by the total counts (6,239 and
20, respectively), and then compared as shown in Fig. 15. The top panel shows
the distributions when all the detected objects are used to select the nearest
neighbors, while the bottom panel shows the result when only the objects with
$H_{160}\leq 25.0$~mag are used. The bottom panel indicates that a fraction of
of our $J_{125}$-dropouts are indeed close to bright, foreground neighbors to
within 1.5$^{''}$, however, this fraction is only $\sim$~40\% (red histogram),
or eight objects in total. The six objects that we visually identified above
are among this population, and the other two objects are C047 and C067, both
are at the border line of being $\sim$~1.5$^{''}$ to their neighbors.
From the distribution of the random positions (black histogram), the
possibility of chance superposition within the same distance is 10.8\% (which
is much larger than the probability of 0.8\% that Bouwens et al. estimated),
and therefore the excess fraction of being close to bright foreground neighbors
is no more than $\sim$~30\%. 

    We should point out that their proximity to foreground neighbors does not
necessarily disqualify these dropouts being legitimate candidates at high
redshifts, because distance to a neighbor is never part of the LBG selection
criterion. It is not likely that they are caused by the noise fluctuations,
because the ``negative image" test does not show that the ``negative dropouts"
are preferentially produced in the regions close to bright objects. To
strengthen this argument, we may examine the similar cases in the
$Y_{105}$-dropout sample where the candidates also have close foreground
neighbors. These $Y_{105}$-dropouts are all detected in two bands as the rest
of the sample, which does not lend support to the suggestion that they are
spurious. It is neither likely that they are caused by the detector defects or
the instrument optics, as such problems would happen to all the three bands and
would affect all dropout samples, and yet no similar cases are seen in the
$z_{850}$-dropout sample. Finally, it is also unlikely that these objects are
physically associated with the neighbors, as none of the galactic stellar
systems that we know would have such colors.

    A possible explanation to these $J_{125}$-dropouts which have close 
foreground neighbors is that they are genuine galaxies at $z\approx 10$
gravitationally lensed by their foreground neighbors. The possibility that a 
significant fraction of galaxies at $z\approx 10$ being gravitationally lensed
by foreground galaxies has been suggested before (e.g., Barkana \& Loeb 2000).
An order-of-magnitude estimate shows that our suggested $z\approx 10$ LF (see
\S 6.3), which is still at the exponential part of the Schechter function 
till 29.4~mag and thus is extremely steep, is broadly consistent with this
interpretation and could explain the observed rate of the close neighbors
because of the magnification bias. Assuming that the lens galaxies are
sigular isothermal spheres (SIS) and that their number density do not evolve,
the lensing probability (before accounting for the magnification bias) at
$z=10$ is to the order of a few per cent (see Kochanek, Schneider \& Wambsganss
2004, eqn.[109] \& [110]). At the separation of $\lesssim 1.5^{''}$ from the
foreground galaxies, the typical magnification factor is $\mu\lesssim 2$. The
surface density of $z\approx 10$ galaxies at 29.00--29.75~mag (i.e., a factor
of 2 lower than the detection threshold of 29.0~mag) as predicted by our LF is
58/arcmin$^2$, which means that $\sim$ 10--11 objects could be magnified to
above our detection threshold if the lensing probability is $\sim$~4\%. Taking
the survey incompleteness into account (see \S 6.3), we would expect 6--8 lensed
$J_{125}$-dropouts in the WFC3 field, which would be in good agreement with the
observed surface density. A detailed lensing analysis is beyond the scope of
this work, and will be presented in a separate paper (Wyithe et al., in prep.). 
The lensing possibility was dismissed by Bouwens et al. on the ground that
there were no highly magnified cases. Their argument again is flawed, because
the probability of $\mu > 2$ is much lower at such separations, and a high
magnification could only be obtained at a much closer distance to the center of
the lens and thus a highly magnified image would not be detected anyway owing
to the blending with the lens itself.

   As we will show in the next section, including or excluding the
$J_{125}$-dropouts that have close foreground neighbors actually does not
statistically change any of our conclusions significantly.

\begin{figure}
\centering
\includegraphics[width=9.0cm]{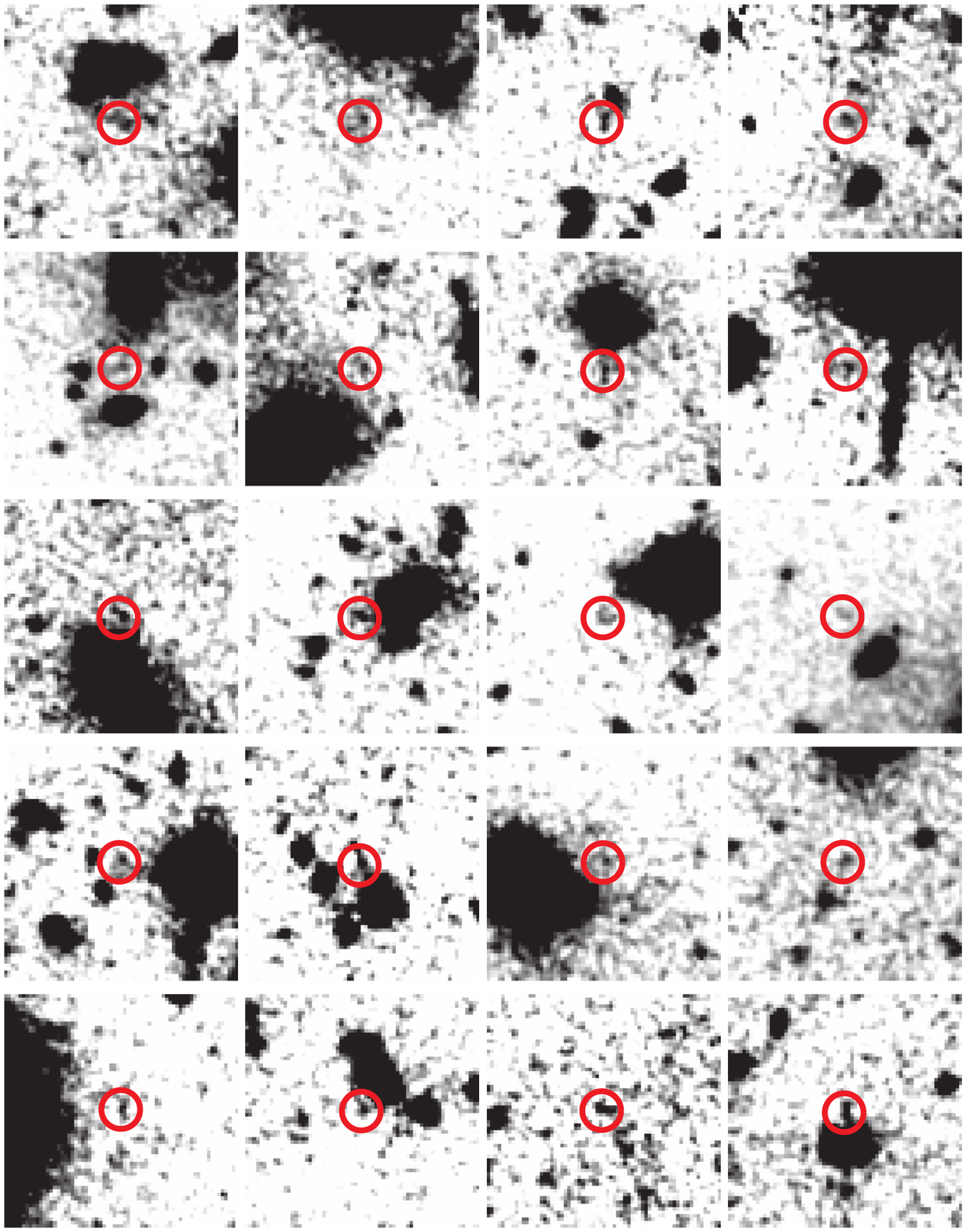}
\caption{$H_{160}$-band cutouts of all the 20 $J_{125}$-dropouts in our sample.
From left to right and then from top to bottom, these objects are
presented in the same order as shown in Table 3 and Fig. 11.
The cutouts are $6^{''}\times 6^{''}$ in size;
north is up and east is left. The red circles are 0.5$^{''}$ in radius, which
clearly show that most of these objects are 0.5${''}$ away from their
bright neighbours, and thus Fig. 7 of Bouwens et al. (2010) is misleading and
incorrect.
}
\end{figure}

\begin{figure}
\centering
\includegraphics[width=9.0cm]{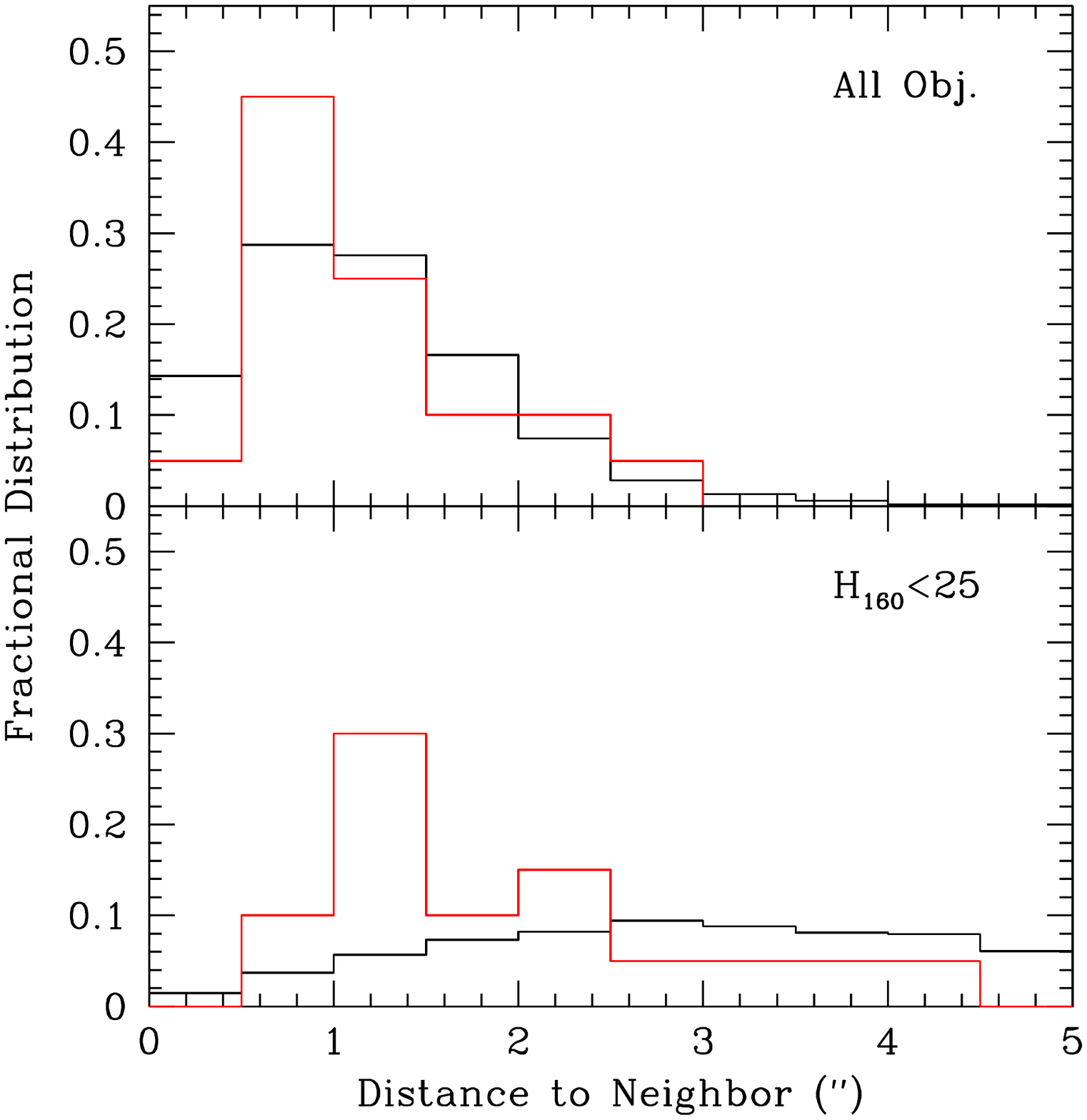}
\caption{Distribution of distances of the $J_{125}$-dropouts to the centroids
of their nearest neighbors (red histograms) as compared to the same distribution
calculated based on 6,239 random sightlines (black histograms). Both histograms
are normalized using the respective total counts. The top panel uses all
objects from the entire $H_{160}$-based catalog, while the bottom panel only
uses the objects with $H_{160}<25.0$~mag for statistics. 
}
\end{figure}

\section{Interpretation and Discussion}

    Our analysis have resulted in 20 $z_{850}$-dropouts, 15 $Y_{105}$-dropouts,
and 20 $J_{125}$-dropouts. While most of our $z_{850}$-dropouts are in common
with those presented by other groups, most of our $Y_{105}$-dropouts and all of
our $J_{125}$-dropouts are new discoveries. As detailed in \S 5, these new
discoveries are the result of a more complete search in our reduced data at the
faint levels. Our search is limited to $\lesssim 29.2$~mag, the limit that
every group has independently adopted.
A small number of candidates reported by other groups are missing from our
samples, and such objects are either outside of our selection field or are at
the borderline of the selection criteria and thus can be explained by our
selection function. 
   In the rest of this section, we discuss the implications of our results.

\subsection{Photometric Redshifts of $z_{850}$-dropouts}

   The dropout selection can only roughly constrain the redshift distribution
of the selected sources by providing a plausible yet rather wide redshift
range to the order of $\Delta z\sim 1$. With the deep multi-band photometry to
the red side of Lyman-break, it is now feasible to further constrain the
redshifts of the $z_{850}$-dropouts in our sample by using the
SED fitting technique to derive their photometric redshifts ($z_{ph}$). As
the vast majority of these sources are much fainter than the limit that can be
reached by the spectroscopic capability of any current instruments, photometric
redshifts will remain as the only redshift estimates for further applications
in the near future. We only carried out this exercises for the
$z_{850}$-dropouts but not for the $Y_{105}$-dropouts, as the latter 
only have useful photometric information in three bands.

   Here we used the SED fitting code developed by HY to derive $z_{ph}$.
The population synthesis models of Bruzual \& Charlot (2003; hereafter BC03)
were used to simulate the fitting templates. We adopted a Simple Stellar 
Population (SSP; i.e., an instantaneous burst) and also a series of star
formation histories (SFH) of exponentially declining star formation rates with
the time scale $\tau$ spanning from 10~Myr to 13~Gyr. The ages ($T$) of
these models range from 1~Myr to 1~Gyr. The initial mass function (IMF) of
Salpeter (Salpeter 1955) was used, with the cut-offs at 0.1 and 100~$M_\odot$.
As we only have a limited number of passbands, we assumed solar metallicity
and zero reddening to reduce the number of free parameters. These assumptions
are justified by the study of Yan et al. (2005), where it is shown that the
properties of galaxies at $z\approx 6$ are consistent with solar metallicity
and very little dust. 
The restframe model spectra were redshifted to $z=5.6$--12.0 at
the step size of $\Delta z=0.1$, and attenuated by the H I absorption along
the line-of-sight using the formalism of Madau et al. (1995). The resulting
spectra were then convolved with the system response curves of the ACS and
the WFC3 IR passbands to generate the model templates for fitting. 

    The SEDs of the $z_{850}$-dropouts were constructed from
the $Y_{105}$-based catalog. We added
0.05~mag to the reported photometric errors in the catalog to account for the
fact that the current WFC3 IR zeropoints are accurate to $\sim 5$\% level. For
the $z_{850}$-dropouts, their fluxes in $B_{435}V_{660}i_{775}$ were set to
zero and thus did not contribute to the fitting. Any non-detections in the
$z_{850}$-band were replaced with the 2-$\sigma$ upper limits measured in the
RMS map at the source locations within a $r=0.2^{''}$ aperture, and a fixed
error of 0.1~mag was assigned to $z_{850}$ in this case.

   The fitting to the model templates were done in the flux domain using the
standard least-square-fit algorithm, with a self-consistency 
constraint that a galaxy should not be older than the age of the universe at the
fit redshift. Four free parameters were involved in this process, namely,
$z_{ph}$, $T$, $\tau$, and stellar mass ($M_*$). The best-fit values of
these parameters are listed in Table 4 for these objects.

   As the WFC3 IR data still only sample the restframe UV wavelengths of these
objects, they cannot break the severe degeneracy between these parameters. For
this reason, the best-fit values listed here, especially those for
$T$, $\tau$ and $M_*$, should only be taken as a general guide line rather than
accurate measurements. The estimate of $z_{ph}$ suffers less from the model
degeneracy, as it mostly depends on estimating the location of the Lyman-break,
which to first order is determined by the line-of-sight H I absorption rather
than the intrinsic properties of galaxies. Fig. 16 shows the histogram of
$z_{ph}$ distributions, which centers at $z\approx 6.9$--7.0 as expected.
We point out that the brightest $z_{850}$-dropout, which YW04b
discovered using the NIC3 data, zdrop-A032, has $z_{ph}=6.9$. This is also
consistent with the reported $z_{ph}=7.0$ in McLure et al. (2009).

\begin{figure}[tbp]
\centering
\includegraphics[width=9.0cm]{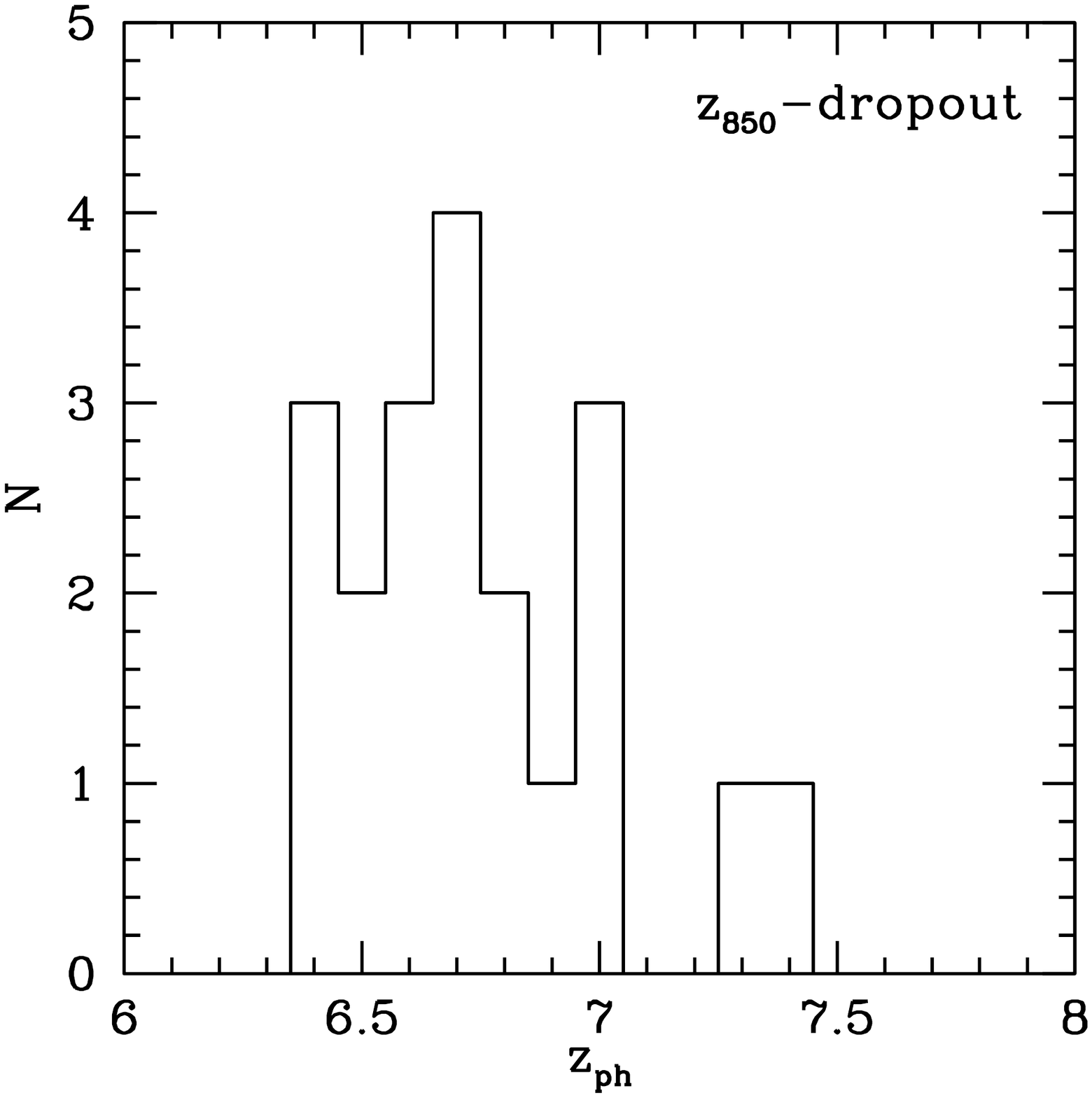}
\caption{Photometry redshift distribution of the $z_{850}$-dropouts,
derived using SED fitting. This distribution is close to the expectation
from the color selection criteria.
}
\end{figure}

\begin{deluxetable}{cccccc}
\tablewidth{0pt}
\tablecolumns{10}
\tabletypesize{\scriptsize}
\tablecaption{Best-fit Model Parameters for $Z\approx 7$ Sample\tablenotemark{a}}
\tablehead{
\colhead{ID} &
\colhead{$\chi^2$} &
\colhead{$z_{ph}$} &
\colhead{$\mathcal{M}_*$/$M_\odot$} &
\colhead{$T$(Myr)} &
\colhead{$\tau$(Gyr)}
}
\startdata
zdrop-A032   &  0.05 &  6.9  & $1.2\times 10^9$ & 100  & 0.5 \\
zdrop-A025   &  0.38 &  7.3  & $1.7\times 10^9$ &  90  & 0.03\\
zdrop-A008   &  0.19 &  7.4  & $1.2\times 10^8$ &  10  & 7.0 \\
zdrop-A060   &  3.17 &  7.0  & $2.0\times 10^8$ &  30  & 0.01 \\
zdrop-A017   &  0.84 &  7.0  & $3.5\times 10^7$ &  10  & 4.5\\
zdrop-A016   &  0.88 &  6.6  & $2.6\times 10^7$ &  10  & 13.0 \\
zdrop-A033   &  4.19 &  6.7  & $1.8\times 10^7$ &   1  & 3.5 \\
zdrop-A014   &  0.54 &  6.7  & $1.9\times 10^9$ & 100  & SSP \\
zdrop-A040   &  0.38 &  6.6  & $2.0\times 10^7$ &   1  & 3.5 \\
zdrop-A044   &  1.35 &  6.8  & $4.8\times 10^9$ & 300  & 0.06 \\
zdrop-A003   &  2.56 &  7.0  & $2.9\times 10^7$ &   1  & 13.0 \\
zdrop-A047   &  0.84 &  6.7  & $1.2\times 10^7$ &   1  &  0.02 \\
zdrop-A057   &  0.10 &  6.4  & $8.0\times 10^6$ &   1  &  0.08 \\
zdrop-A046   &  4.59 &  6.8  & $2.4\times 10^7$ &  10  &  0.8 \\
zdrop-A053   &  0.08 &  6.7  & $4.9\times 10^7$ &  40  &  2.0 \\
zdrop-A056   &  5.22 &  6.5  & $9.7\times 10^6$ &   1  &  7.0 \\
zdrop-A052   &  1.74 &  6.5  & $1.0\times 10^7$ &   1  &  0.9 \\
zdrop-A055   &  0.48 &  6.6  & $1.2\times 10^7$ &   1  &  0.07 \\
zdrop-A062   &  1.17 &  6.4  & $8.1\times 10^6$ &   1  &  8.0 \\
zdrop-A065   &  0.59 &  6.4  & $8.2\times 10^6$ &   1  &  1.5 \\

\enddata
\tablenotetext{a.}{Fitting templates derived using BC03 models, Salpeter IMF, solar metallicity, and zero reddening.}
\end{deluxetable}

\subsection{Stellar Masses of Galaxies at $z\approx 7$ and Beyond}

   Except two previously studied sources, no other objects in our three dropout
samples show convincing case of detection in the deep GOODS IRAC images.
These two sources are the two brightest $z_{850}$-dropouts, zdrop-A032 and A025.
In the IRAC images, their locations are unfortunately very close to other
unrelated objects nearby, and the extraction of their fluxes in IRAC passbands
is non-trivial. Nevertheless, their stellar masses have been derived by
Labb\'{e} et al. (2006), using the deblended IRAC photometry as the major
constrain on
their SEDs in the rest-frame optical. Depending on the models used, they
derived stellar masses of a few $\times 10^9 M_\odot$ at $z_{ph}\approx 6.8$
and 7.3 for zdrop-A032 (their ID 1147) and A025 (their ID 963), respectively.
Interestingly, these values are not inconsistent with the best-fit values that
we have obtained (see Table 4) using only the rest-frame UV flux measurements.
There are a couple of more cases in Table 4 where the inferred stellar masses
are at $\sim 10^9 M_\odot$ level, but none of them are significantly higher.
This is in contrast to the $i_{775}$-dropout
sample in the HUDF, where three $z\approx 6$ objects are significantly
detected in IRAC $3.6$ and $4.5$~$\mu$m images and have stellar masses to the
order of a few $\times 10^{10}M_\odot$ (Yan et al. 2005; Eyles et al. 2005). 

   As an effort to further constrain the average stellar mass of galaxies at
$z\approx 7$ and beyond, we stacked these objects in both IRAC 3.6~$\mu$m and
4.5~$\mu$m channels. Sources that are not obviously blended with foreground
objects (A032 and A025 are deemed to be blended) were selected, and their
images in IRAC 3.6~$\mu$m and 4.5~$\mu$m were median-combined separately in
the usual way. Objects that are close pairs were considered single
objects (as they would not be separable by IRAC) and were only counted once.
We first did this exercise for all three samples separately,
and then also did the same for the merged $z_{850}$- and $Y_{105}$-dropout
sample. We did not see any detection in any of these cases. As no positive
signal was detected, and the number of stacked objects are still limited,
we are not able to derive any constrain stronger than a general statement
that the majority galaxies at $z\approx 7$ and beyond have stellar mass
upper limits to the order of $10^9 M_\odot$.

    Encouraged by the agreement between our results and those of Labb\'{e}
et al.'s
for zdrop-A032 and A025, we added up the stellar masses listed in Table 4 to
derive the global stellar mass density at $z\approx 7$. This crude estimate
gives $5.3\times 10^5 M_\odot$~Mpc$^{-3}$, which is a factor of
10--12 smaller than that at $z\approx 6$.

\begin{figure}
\centering
\includegraphics[width=13.0cm]{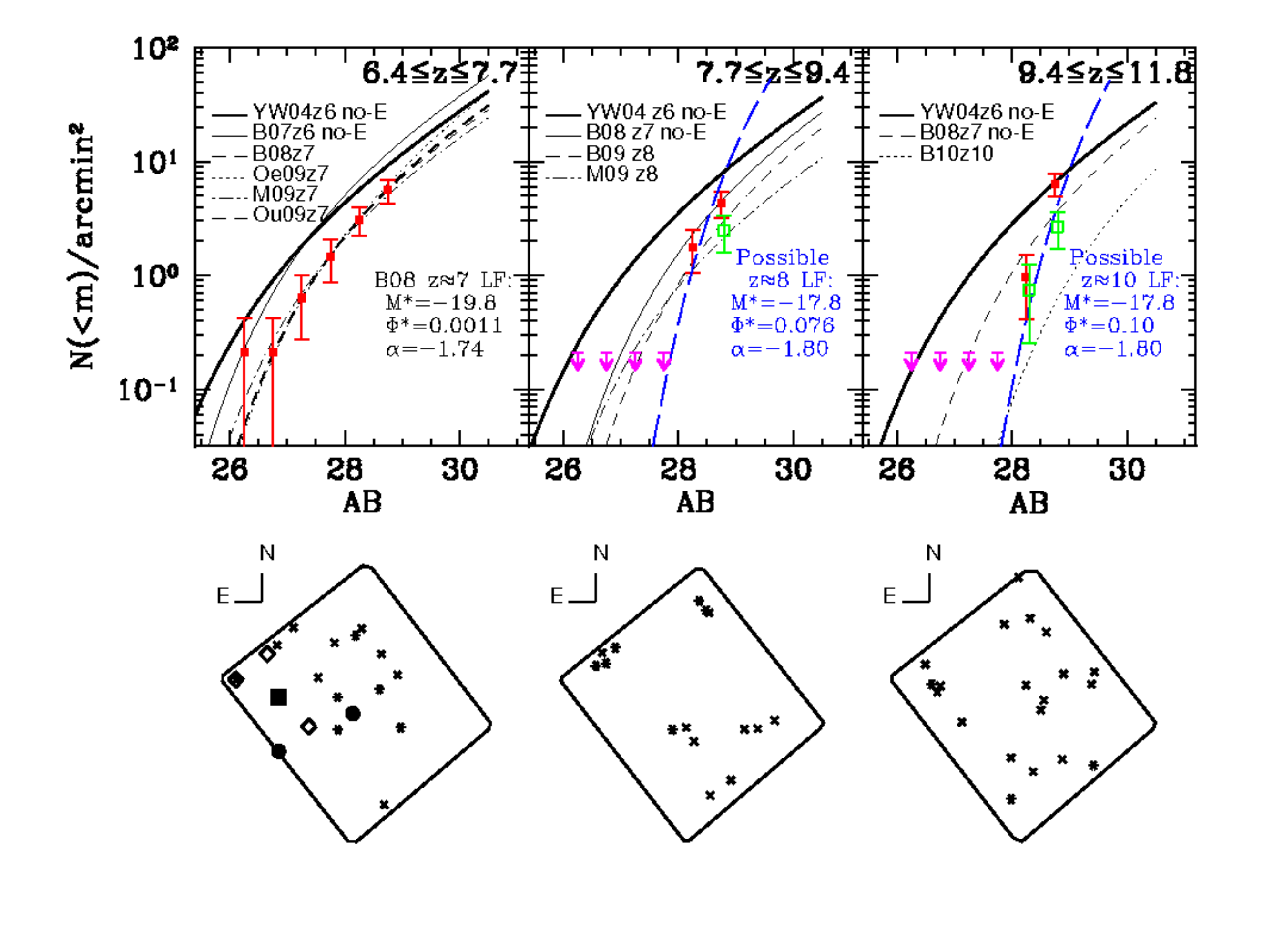}
\caption{(top) Cumulative surface densities of the dropouts, compared to the
predictions of various LF estimates. The data points have all been corrected
for the survey incompleteness, and the error bars indicate 1~$\sigma$ Poisson
noise. The green point in the middle panel includes only the $Y_{105}$-dropouts
that satisfy $Y_{105}-J_{125}\geq 0.8$~mag when using the 2~$\sigma$ limits in
$J_{125}$-band. The green points in the right panel have excluded the
$J_{125}$-dropouts that might be affected by the gravitational lensing of
their foreground neighbors, and have taken the possible
contamination due to the noise fluctuation into account.
 At $6.4\leq z\leq 7.7$, the observations agree with various
$z\approx 7$ LF estimates. At $7.7\leq z\leq 9.4$ and $9.4\leq z\leq 11.8$,
the observations show a striking feature that there is no dropout at
$m\leq 28.0$~mag. In both cases, the firm non-detections at 
$m\leq 28.0$~mag (purple upper limits) and the firm detections above it
(especially in the $28.0<m\leq 28.5$~mag bin) suggest that the underlying LF
must be steeply rising towards lower luminosity. The stepness is consistent
with the exponential part of the Schechter function, and we suggest possible
LFs accordingly. Our tentative LFs are shown as the dashed blue curves. In
particular, the LF at the $z\approx 10$ case is estimated {\it after} excluding
the close-neighbor cases in the $J_{125}$-dropout sample.
(bottom) One potential source of uncertainty that can affect our interpretation
is the effect of spatial clustering. While currently we are not able to
derive useful constraint from the data, we show the spatial locations of the
dropouts in these schematic figures (one for each sample), and point out that
some apparent clustering signature does present. The symbols are coded
according to their magnitudes:
filled square --- $m\leq 26.5$, filled circle --- $27.0< m\leq 27.5$, 
diamond --- $27.5< m\leq 28.0$, stars --- $28.0< m\leq 28.5$, 
cross --- $28.5< m\leq 29.0$. We do not have any candidates at 
$26.5\leq m<27.0$ at any of these redshifts.
}
\end{figure}

\subsection{Constraint to LF of Galaxies at Very High Redshifts}

   Our large dropout sample has offered a new opportunity to constrain
the LF of galaxies at $z\approx 7$--10. Fig. 17 shows our data points in red,
derived based on the dropouts listed in Table 1, 2 and 3, with error bars
representing the Poisson noise. Three candidates that are formally (though
only slightly) fainter than 29.0~mag (i.e., Ydrop-SD24, Jdrop-C056 and
Jdrop-C034) are excluded from the statistics. For ease of connection to
observations, the data are presented in form of cumulative surface density per
arcmin$^2$ as a function of apparent magnitude in bin size of 
$\Delta m=0.5$~mag, i.e.,
$N(\leq 26.5, 27.0, 27.5, 28.0, 28.5, 29.0)$, plotted at the center of each
magnitude bin, i.e, at $m=(26.25, 26.75, 27.25, 27.75, 28.25, 28.75)$~mag.
To be conservative, we also consider the possibility that our samples could be
significantly contaminated at $z\approx 8$ and 10.
In case of $Y_{105}$-dropouts, the green symbol in the 28.75~mag bin represents
the surface density after excluding the objects that do not strictly
satisfy $Y_{105}-J_{125}\geq 0.8$~mag if using the 2 $\sigma$ $Y_{105}$-band
limits, which are all at $J_{125}\geq 28.5$~mag (see Table 2 and \S 5.2).
In case of $J_{125}$-dropouts, the green symbols represent the results after
excluding the objects that have close neighbors (see \S 5.4) and after
taking into account the possible contamination due to noise fluctuation
(see \S 5.3). These green symbols are offset slightly in magnitude
for clarity. The cumulative surface densities predicted by various LFs are also
shown for comparison. In all cases we use the non-evolving LF of YW04b at
$\approx 6$ as the fiducial (hereafter YW04z6LF), 
which has the Schechter function parameters of 
$M^*=-21.03$~mag, $\Phi^*=4.6\times 10^{-4}$~Mpc$^{-3}$, and $\alpha=-1.8$. 

   The data points presented here have all been corrected for the survey
incompleteness.
We determine the incompleteness correction following the method used in YW04.
For each dropout, a $15\times 15$ pixel image stamp centered at its location in
the discovery band was cut out and then randomly distributed to $\sim 100$
positions in the discovery image. For the $z_{850}$-dropouts, this was done
in the $Y_{105}$-band mosaic. For the $J_{125}$-based and the $JH$-based
$Y_{105}$-dropouts, this was done in the $J_{125}$-band mosaic and the
$JH$-mosaic, respectively. For the $J_{125}$-dropouts, this was done in the
$H_{160}$-band mosaic. The sections of the discovery science images at these
random locations were relpaced by the image cutouts, but the RMS maps were not
altered. This approach was to simulate objects of the same properties
(such as the brightness and the morphology) as the dropout under study
and to put them to different noise background (i.e., corresponding to
different locations in the RMS maps). SExtractor was run on these new images
following the same procedures as in \S 4, and we applied the same $S/N\geq 3$
thresholds to recover the simulated ``dropouts''. The rate of recovery was
taken as the completeness of this particular kind of dropouts, and the average
recovery rate of all the dropouts within a magnitude bin was taken as the
completeness of this bin. For the $z_{850}$-dropout sample, it is complete in
the $Y_{105}=27.25$~mag bin and brighter; the incompleteness correction is a 
factor of 1.3, 1.3, 2.0 for the differential counts in the 27.75, 28.25 and
28.75~mag bins, respectively. For the $Y_{105}$-dropout sample, the
incompleteness correction is a factor of 1.3 and 1.6 in the $J_{125}=28.25$ and
28.75~mag bins, respectively. For the $J_{125}$-dropout sample, the correction
is a factor of 1.5 and 1.7 in the $H_{160}=28.25$ and 28.75~mag bins,
respectively.

   At $6.4\lesssim z\lesssim 7.7$ (top left in Fig. 17), our data agree quite
well with the LF estimate of B08 at $z\approx 7$ (hereafter B08z7LF) and the
new estimates of Ouchi et al. (2009; Ou09z7), Oesch et al. (2009; Oe09z7) and 
McLure et al. (2009; M09z7) at $m\geq 27.5$~mag. The agreement at 
$m\leq 27.0$~mag with any LFs, however, is not satisfactory. The data points in
these two magnitude bins
are the result of one single detection at the bright end, namely zdrop-A032.
Therefore, the discrepancy could possibly be due to the combined effect of
small number statistics and ``cosmic variance''.
Indeed, the spatial distribution of these dropouts shows apparent clustering,
as indicated in the schematic plot at the bottom right panel of Fig. 17.
A wider survey is necessary to properly address the bright-end behavior. 

   At $7.5\lesssim z\lesssim 9.4$ (top middle in Fig. 17), our data show a
striking feature that there is no $Y_{105}$-dropout at $J_{125}\leq 28.0$~mag.
This is noted in Bouwens et al. (2009) as well, and here we see it at higher
significance with our larger sample. Given the large uncertainty in the
current measurement, especially at the faintest level (see the red and the
green symbols), both of the LFs suggested by Bouwens et al. (2009) and McLure
et al. (2009) are consistent with our data points. However, neither of them 
seem to be consistent with the upper limits at $J_{125}\leq 28.0$~mag: those
LFs would predict 2--3 objects to this brightness level, where the current data
are essentially complete (see Fig. 3). To satisfy both the data points and the
upper limits, it seems that the underlying LF must be steeply rising towards
faint luminosity. If the LF still takes the form of the Schechter function,
the current data, while having reached $\sim 29$~mag ($M\sim -18.3$~mag),
should still be sampling the exponential part of the LF. Motivated by this
argument, we propose a possible LF that has the following Schechter function
parameters: $M^*=-17.8$~mag, $\alpha=-1.8$, $\Phi^*=0.076$~Mpc$^{-3}$.
The cumulative surface density predicted by this LF, shown as the blue curve
in the figure, is consistent with {\it both} the data points (red points)
{\it and} the upper limits. Taking it at the face value, this LF implies a 
dimming of $2.3$~mag in $M^*$ and a striking increase by a factor of 
$\sim 17\times$ in $\Phi^*$ as compared to the LF at $z\approx 7$. 

   And it seems that we could draw a similar conclusion on the LF at 
$9.4\leq z\leq 11.8$ (top right in Fig. 17). The inferred cumulative surface
density from our $J_{125}$-dropout sample is characterized by the non-detection
at $H_{160}\leq 28.0$~mag and the steep increase in the last two bins, and we
propose the following Schechter parameters, tuned to better fit the green
symbols (i.e., excluding the objects with close foreground neighbors and taking
into account the possible contaimation due to the noise fluctuation):
$M^*=-17.8$~mag, $\alpha=-1.8$,
$\Phi^*=0.10$~Mpc$^{-3}$. The predicted cumulative surface density from this LF
is shown as the blue curve in the figure. Note that $\Phi^*$ increases by a
factor of $\sim 90$ as compared to the LF at $z\approx 7$. 

  For convenience, we list the LF parameters from this study together with
those of other LFs in Table 5. Again, a direct comparison of those seems to
suggest a sudden change in behavior of the galaxy LF at $z\approx 8$. While our
proposed LFs at $z\approx 8$ and 10 are not yet demanded by the current
observations, they are {\it allowed} by the data, and seem to fit better than 
other alternatives.

\begin{deluxetable}{rccc}
\tablewidth{0pt}
\tablecolumns{10}
\tabletypesize{\scriptsize}
\tablecaption{Schechter Function Parameters of Various LFs Disscused}
\tablehead{
\colhead{LF Source\tablenotemark{a}} &
\colhead{$M^*$~(mag)} &
\colhead{$\Phi^*$~(Mpc$^{-3}$mag$^{-1}$)} &
\colhead{$\alpha$}
}
\startdata

 YW04z6LF, $z\approx 6$      & $-21.03$ & 4.6$\times 10^{-4}$ & $-1.80$ \\
  B07z6LF, $z\approx 6$      & $-20.24$ & 1.4$\times 10^{-3}$ & $-1.74$ \\
  B08z7LF, $z\approx 7$      & $-19.80$ & 1.1$\times 10^{-3}$ & $-1.74$ \\
 Ou09z7LF, $z\approx 7$      & $-19.90$ & 1.1$\times 10^{-3}$ & $-1.70$ \\
  Oe09z7LF, $z\approx 7$      & $-19.80$ & 1.1$\times 10^{-3}$ & $-1.86$ \\
  M09z7LF, $z\approx 7$      & $-20.11$ & 7.0$\times 10^{-4}$ & $-1.72$ \\
 this work, $z\approx 8$     & $-17.80$ & 7.6$\times 10^{-2}$ & $-1.80$ \\
 this work, $z\approx 10$    & $-17.80$ & 1.0$\times 10^{-1}$ & $-1.80$ \\

\enddata
\tablenotetext{a.}{Ref: YW04z6LF -- Yan \& Windhorst (2004b); 
B07z6LF -- Bouwens et al. (2007); 
B08z7LF -- Bouwens et al. (2008);
Ou09z7LF -- Ouchi et al. (2009);
Oe09z7LF -- Oesch et al. (2009);
M09z7LF -- McLure et al. (2009)
}
\end{deluxetable}

\begin{figure}
\centering
\includegraphics[width=9.0cm]{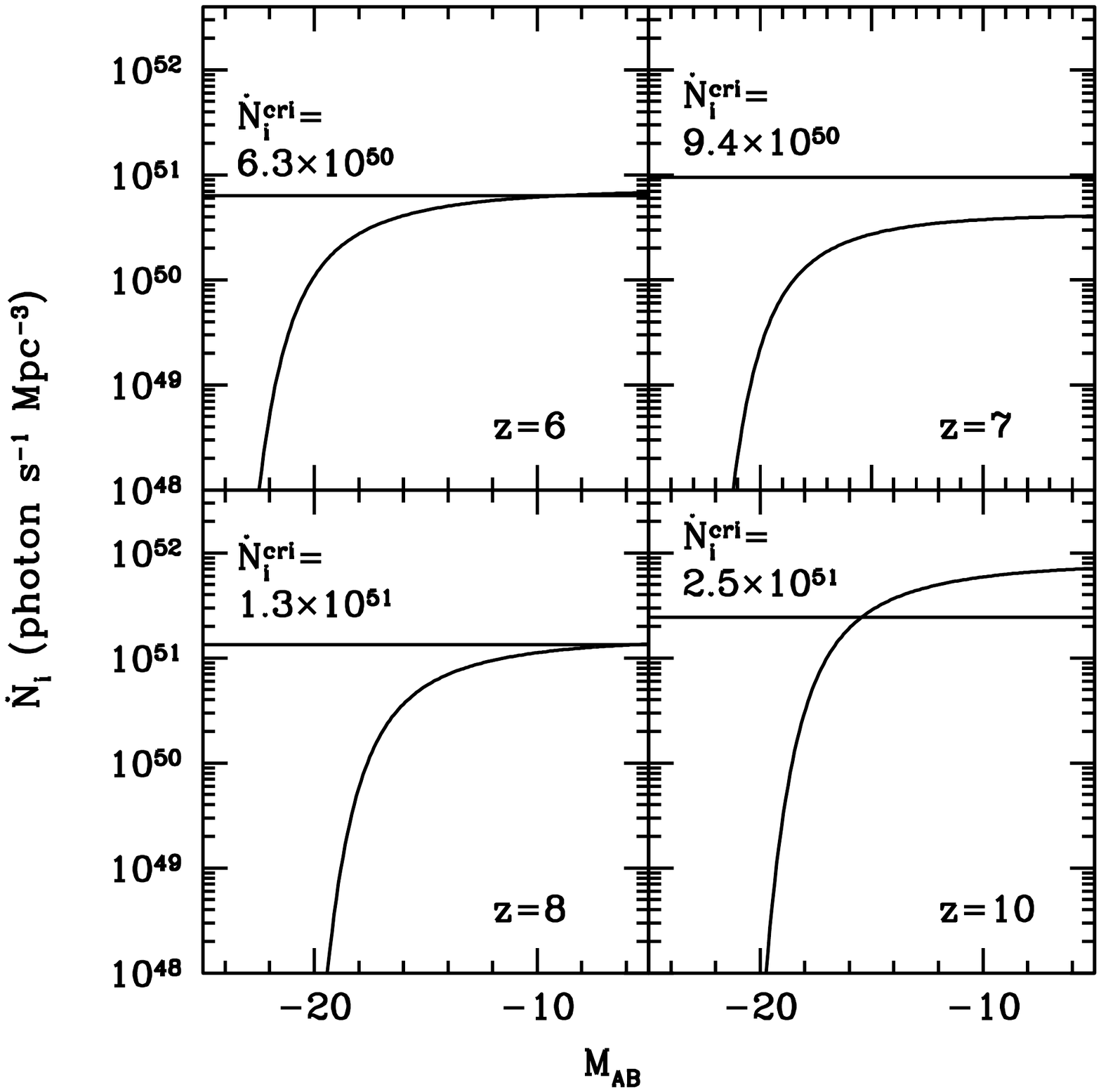}
\caption{Ionizing photon production rate density from galaxies
($\dot{N}_{i}$; the curves) 
compared to the critical value ($\dot{N}_{i}^{cri}$; the horizontal
lines) that is required to keep the universe completely ionized at various
redshifts. As presented in YW04a, galaxies can account for the entire ionizing
photon budget at $z\approx 6$ if their LF has a steep faint-end slope $\alpha$.
Assuming galaxies have similar intrinsic properties, the calculated 
$\dot{N}_{i}$ at $z\approx 7$ using B08z7LF (which has a steep $\alpha$)
could be $\sim 40$--50\% of $\dot{N}_{i}^{cri}$, which might be
desirable because one would expect the neutral fraction of hydrogen at
$z\approx 7$ is still $>>0$. If we apply the LFs that we tentatively propose
for $z\approx 8$ and 10 (see Table 5), the implied $\dot{N}_{i}$ would
cross over $\dot{N}_{i}^{cri}$ if the LFs do not cut-off and the
reionization would be completed too early.
}
\end{figure}

\begin{figure}
\centering
\includegraphics[width=9.0cm]{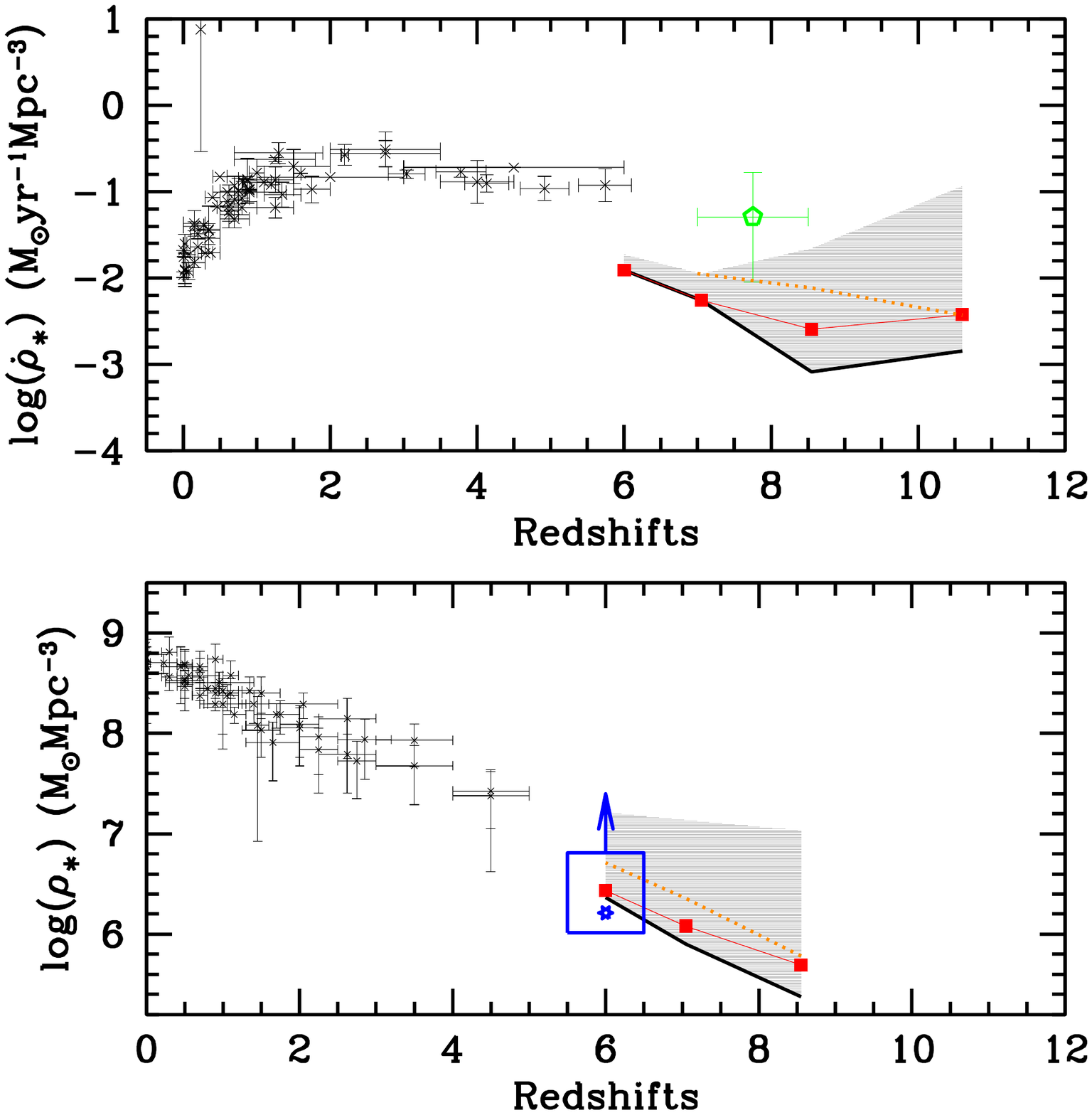}
\caption{(top) Evolution of the global SFR density (GSFRD; $\dot{\rho_{\ast}}$)
from $z\approx 10$. The values at $z<6$ are taken from Hopkins \& Beacom (2006).
The grey region reflects the uncertainty in the current estimates at
$z\gtrsim 6$. The red squares at $z=6.0$ and 7.0 are obtained by integrating
YW04z6LF and B08z7LF to the dropout search limit of $m=30.0$ (for $z\approx 6$)
and 29.0~mag (for $z\approx 7$), respectively. The red squares at $z=8.6$ and
10.6 are derived by adding the contribution from our $Y_{105}$- and
$J_{125}$-dropouts, respectively, and have been applied the proper correction
for the imcompleteness. The bottom boundary of this region (indicated by the
thick black line) shows the GSFRD derived based on the conservative estimates
of the dropout number densities at $z\approx 8$ and 10 (but still having the
incompleteness correction applied). For $z\approx 8$, this estimate only takes
into account the $Y_{105}$-dropouts that satisfy
$Y_{105}-J_{125}\geq 0.8$~mag when using the 2~$\sigma$ limits in $Y_{105}$,
and is calculated using the 1~$\sigma$ lower limit (due to Poisson noise) of
the $Y_{105}$-dropout surface density thus inferred.
For $z\approx 10$, this estimate is obtained by statistically subtracting the
possible contamination due to the noise fluctuation, and is also calculated
using the 1~$\sigma$ lower limit of the surface density thus inferred.
The top boundary of the grey region is obtained
by integrating the LFs to the fiducial limit of $M=-15$~mag. For $z=6.0$ and
7.0, YW04z6LF and B08z7LF are used, respectively. For $z=8.6$ and 10.6,
our proposed LFs at $z\approx 8$ and 10 are used, respectively. The orange
dotted line represents the results by integrating the LFs at $z\approx 7$, 8
and 10 of Bouwens et al. (2007, 2009, 2010) to $M=-15$~mag. The green
symbol is the GSFRD at $z\approx 8$ as derived in Kistler et al. (2009)
based on GRB 090423 at $z\approx 8.1$.
(bottom) Evolution of the global stellar mass density (GSMD; $\rho_{\ast}$) from
$z\approx 10$. The values at $z<6$ are taken from Wilkins et al. (2008). 
The grey region here maps the grey region in the top panel (including the
symbols), and is obtained by
integrating the GSFRD over time (assuming zero stellar mass density at $z=10$).
The blue star at $z=6.0$ and the surrounding box, taken from Yan et al. (2006),
represent the GSMD estimate at this redshift and the associated uncertainty,
which should be taken as a strict lower limit because only detected galaxies
were used. It is likely that the vast majority of the stellar masses assembled
over the reionization epoch is still undetected at $z\approx 6$.
}
\end{figure}
\subsection{Global Star Formation Rate Density and Stellar Mass Density}

\subsection{Implications for Cosmic Hydrogen Reionization}

  As $z\approx 8$--10 extend well {\it into} the cosmic reionization epoch, and
our LFs sample an area in the parameter space that has not yet been explored,
here we discuss their implication for the reionization.

   A question that has been intensively studied is whether the
star-forming galaxy population can be the source of cosmic hydrogen
reionization, or at least be an important part of it. To answer this question,
one can calculate the integrated production rate density of photons at 
$\lambda< 912$~\AA\, from galaxies ($\dot{N}_{i}$), and then compare to
the critical ionizing photon emission rate density 
($\dot{N}_{i}^{cri}$) that is necessary to balance the combination
rate. Madau, Haardt \& Rees (1999; MHR) have given a recipe (their eqn.[26])
to calculate
$\dot{N}_{i}^{cri}$ at arbitrary redshifts, with the only free
parameter being the hydrogen clumping factor ($C$).
Calculating $\dot{N}_{i}$ is done by integrating the LF and then
multiplying by a factor that describes the product of the number of Lyman
photons produced per unit galaxy (which in principle can be inferred from the
UV slope of typical LBGs) and the fraction of those that can escape the
galaxy ($f_{esc}$). Another approach is to compare to the critical SFR density
($\dot\rho_{\ast}^{cri}$) that is derived from $\dot{N}_{i}^{cri}$ 
using galaxy population synthesis model (given in MHR eqn.[27]). To do this,
one needs to know the SFR of a typical galaxy, and this is almost always done
by using the conversion between $L_{UV}$ and SFR
as given in Madau, Pozzetti \& Dickinson (1998; their eqn.[2]). A full
discussion of the application of these two approaches is beyond the scope of
this paper; here we only point out that (1) currently the results from these
two methods do not necessarily agree, (2) comparing to 
$\dot\rho_{\ast}^{cri}$ implicitly assumes a particular population synthesis
model and the particular IMF that this model takes, and (3) comparing to
$\dot{N}_{i}^{cri}$ only depends on the choice of the SED shape
at $\lambda< 912$~\AA\, but does not depend on any model.

    YW04a studied this problem at $z\approx 6$ using this first approach, and
proposed a steep LF faint-end slope of $\alpha < -1.6$ in order to bring 
$\dot{N}_{i}$ to meet $\dot{N}_{i}^{cri}$. YW04b soon confirmed
that $\alpha\lesssim -1.8$ using the HUDF data, and pointed out that 
star-forming galaxies alone could have contributed most of the reionizing
photon background. Since then, the very steep LF faint-end slope at 
$z\approx 6$ has been confirmed by other studies (e.g., Bouwens et al. 2006;
B07), and it has been widely accepted that star-forming galaxies, especially
those at the faint-end, must have played an important role in the reionization.
YW04a adopted $f_\nu\propto \nu^{-1.8}$ at $\lambda < 912$\AA, which would
allow the YW04z6LF to produce $\dot{N}_{i} = \dot{N}_{i}^{cri}$
at $z=6$ if integrating the LF to
$M=-15.7$~mag and assuming $f_{esc}=0.12$ and $C=30$. Both $f_{esc}$ and $C$
are highly uncertain. While the recent simulation of Pawlik et al. (2009)
suggests a much lower clumping factor of $C=6$, the studies of $f_{esc}$
seem to point to a small value of $f_{esc}\lesssim 0.02$--0.05
(e.g., Siana et al. 2007; but see also Shapley et al. 2006).
The combination of the two would still make 
$\dot{N}_{i}$ meets $\dot{N}_{i}^{cri}$, only that the
crossover now is at $M=-9.0$~mag. 
Adopting B07z6LF would give a similar answer, with the only
difference being that this LF only needs to be integrated to $M=-13.3$~mag
to have the crossover.

   Here we consider how our new results could shed light to the source of
reionization problem in the following context. Assuming that the SED
power law index adopted by YW04a is appropriate such that the galaxy population
alone can indeed sustain the ionizing background at $z\approx 6$, would the
galaxies at $z\approx 7$--10 produce the right amount of ionizing photons 
if these galaxies have the same properties as those at $z\approx 6$? Simply
speaking, now we hope that 
$\dot{N}_{i} \lesssim \dot{N}_{i}^{cri}$,
because we do want a large number of ionizing photons such that the 
reionization can happen, and at the same time we do not want too many
ionizing photons such that the reionization would be ended too early and there
would be no neutral hydrogen left to create the Gunn-Peterson troughs that
have been observed at $z\approx 6.5$. YW04a has investigated this problem and
pointed out that one solution
is to have the Schechter LF broken down at some certain minimum luminosities to
prevent this from happening. As we now have hints that the galaxy LF at
$z\approx 8$ and beyond might imply an unexpectedly large number of faint
galaxies, it is prudent to investigate this problem further.

   Our results are summarized in Fig. 18, assuming $C=6$ and $f_{esc}=0.02$.
For simplicity, we quote $\dot{N}_{i}$ in units of 
$10^{51}$~photons~s$^{-1}$~Mpc$^{-3}$.
At $z=6$, $\dot{N}_{i}$ is calculated using YW04z6LF, and it crosses
$\dot{N}_{i}^{cri}$ at around $M=-9.0$~mag as 
mentioned before. At $z=7$, $\dot{N}_{i}$ is calculated using B08z7LF,
which is consistent with the new observations as discussed in the previous
section. In this case, $\dot{N}_{i}$ asymptotically approaches
$0.42$, which is about a factor of $2.2\times$ less than
$\dot{N}_{i}^{cri}$. This might be somewhat too low, but it does
satisfy the requirement that the ionizing photons are not overproduced.
At $z=8$, $\dot{N}_{i}$ calculated using the 
LF parameters quoted in Table 5 does not cross over
$\dot{N}_{i}^{cri}$ until $M=-6.7$~mag. It is not clear if the
Schechter function still holds at such lower luminosity levels; if it does,
in order {\it not} to reionize the universe completely at $z=8$, it
seems that the contribution of ionizing photons from objects at $M>-6.7$~mag
should be cut-off, for example, by invoking $f_{esc}=0$ for
these galaxies. One way to achieve the cut-off is to completely shut down the
star formation. As very low mass halos are not capable of cooling down to a
sufficiently low temperature to form stars, such a cut-off might not be
surprising.  At $z=10$, the cross-over happens at a much brighter level
of $M=-15.4$~mag, and the similar reasoning suggests that
$f_{esc}=0$ is required for galaxies fainter than this threshold. This would
mean that most of the ionizing photons at these stages are from a number of the
brightest objects, and this does seem to be consistent with a sudden on-set of 
reionization. Of course, the above numbers should only be taken as a guide,
because we do not yet have any definite knowledge about $C$ and $f_{esc}$, nor
do we know the LF very well. Nevertheless, this very simple exercise could be
potentially very useful.

   If we adopt the conversion between $L_{UV}$ and SFR as 
$L_{UV}=8.0\times 10^{27} \times SFR$ (MPD eqn.[2]) for the Salpeter IMF, we can
easily calculate the global SFR density (GSFRD; $\dot{\rho_{\ast}}$) at
$z\gtrsim 7$. The results are shown in the top panel of Fig. 19 for
$z\approx 7$, 8, and 10, together with the GSFRD at $z\approx 6$ for
comparison. We do not include the correction for dust extinction at these four
redshifts, as it is not well constrained at the moment. To see the overall
evolution trend, the GSFRD at lower redshifts (Hopkins \& Beacom 2006) are also
shown.

   In Fig. 19, the red squares at $z=6.0$ and 7.0 are obtained by integrating
the LFs in the luminosity regimes that have been directly probed by the existing
observations. While the LFs derived by various groups are different, they all
agree with the direct observations down to the survey limits. Therefore, by 
integrating to the survey limits only, one can take the derived values as the
lower bounds of the GSFRD. Specifically, the one at $z=6.0$ is obtained by
integrating YW04z6LF to 30.0~mag, which is the limit of the $i_{775}$-dropout
search in the ACS HUDF. Similarly, the one at $z=7.0$ is derived by integrating
B08z7LF to 29.0~mag, which is the limit of the $z_{850}$-dropout search in the
current WFC3 HUDF. These values are 
$\dot{\rho_{\ast}}=(12.33, 5.50)\times 10^{-3}$~$M_\odot$~yr$^{-1}$~Mpc$^{-3}$
at $z=(6.0, 7.0)$, respectively.

   The red squares at $z=8.6$ and 10.6 are obtained by adding the contribution
from the observed dropouts after applying proper corrections for the survey
incompleteness. In other words, they correspond to the red symbols in Fig. 12.
These values are 
$\dot{\rho_{\ast}}=(2.54, 3.77)\times 10^{-3}$~$M_\odot$~yr$^{-1}$~Mpc$^{-3}$
at $z=(8.6, 10.6)$, respectively.

   As the observations at $z\approx 8$ and 10 still have large uncertainties,
it is useful to consider the lower limits of the GSFRD at these redshifts. 
These are shown in Fig. 19 as the thick solid line that outlines the bottom
of the grey region. The lower limit at $z\approx 8$ is obtained by including
only the most robust $Y_{105}$-dropouts that satisfy 
$Y_{105}-J_{125}\geq 0.8$~mag when using the 2~$\sigma$ upper limit in  
$J_{125}$ (i.e., the objects in the first part of in Table 2), and is
calculated using the 1~$\sigma$ lower limit (due to Poisson noise) of the
surface density thus derived. 
The lower limit at $z\approx 10$, on the other hand, is obtained by statiscally
subtracting the contribution from the possible contamination from the noise
fluctuation, and is also calculated using the 1~$\sigma$ lower limit of the
surface density thus inferred. Specifically, these values are
$\dot{\rho_{\ast}}=(0.82, 1.42)\times 10^{-3}$~$M_\odot$~yr$^{-1}$~Mpc$^{-3}$
at $z=(8.6, 10.6)$, respectively.

   We should also consider how high the GSFRD could possibly be. This would
require extrapolate the LFs and then integrate, and is somewhat uncertain in
the sense that the limit to which the extrapolation (and hence the integration)
should stop is arbitrary. 
For the sake of the argument, here we choose to stop
at $M=-15.0$~mag, or $0.01\times L^*(z=3)$. At $z\approx 10$, this limit
corresponds to $m\sim 32.7$, which would be difficult
to reach even with the JWST. The results of this exercise are
$\dot{\rho_{\ast}}=(0.019, 0.011, 0.022, 0.116)$~$M_\odot$~yr$^{-1}$~Mpc$^{-3}$
at $z=(6.0, 7.0, 8.6, 10.6)$, respectively. 
The values at $z=6.0$ and 7.0 are
based on YW04z6LF and B08z7LF, respectively, while the values at $z=8.6$ and
10.6 are based on the LFs that we propose in \S 6.3.
These values are shown as the upper envelope of the grey region in Fig. 19.
If a fainter integration limit is adopted, this upper envelope will be higher.
For comparison, the orange dotted line shows the GSFRD calculated based on the 
$z\approx 8$ and 10 LFs of Bouwens et al. (2009, 2010), integrated to the same
limit of $M=-15.0$~mag.

   This grey region reflects the uncertainties in the current GSFRD
measurement at $z\gtrsim 7$. The major source of uncertainty is in the LF
estimates at $z\gtrsim 8$, which is not likely to be improved until
significantly deeper and wider data are available. Nevertheless, it is worth
pointing out that if our proposed LFs at $z\approx 8$ and 10 are correct, the
evolution of the GSFRD could be very different from what one would extrapolate
from the trend seen at lower redshifts. As the upper envelope of the grey area
implies, the GSFRD could start at early time at a very high value, decline to a
valley at $z\approx 7$, and then increase again towards $z\approx 6$. While the
suggestion of this trend is still only tentative, such a behavior of the GSFRD
in the early time seems to fit naturally into the picture of the reionization.
A very high GSFRD in the early universe has also been suggested by Kistler
et al. (2009) based on Gamma-Ray Bursts (GRBs) at high redshifts, in particular
GRB 090423 at $z\approx 8.1$ (Salvaterra et al. 2009; Tanvir et al. 2009). The
green pentagon in the upper panel of Fig. 19 shows the GSFRD at $z\approx 8$
as derived in Kistler et al. (2009). If our LF at $z\approx 8$ is integrated to
$M\approx -8.5$~mag, the inferred GSFRD will match that of Kister et al.
Using the LF of Bouwens et al. (2009) or McLure et al. (2009), however, the
inferred GSFRD will be a factor of 5 too low even when integrating to zero
luminosity.

    From the GSFRD we can obtain the {\it expected} global stellar mass 
densities (GSMD; $\rho_{\ast}$) by a straightforward integration over time.
Assuming that the universe started from zero stellar mass at $z\approx 10$,
the GSMD thus calculated are shown as the grey region in the bottom panel of
Fig. 19, which maps the grey GSFRD region in the top panel. The lower boundary
corresponds to
$\rho_{\ast}=(0.23, 0.08, 0.02)\times 10^7$~$M_\odot$~Mpc$^{-3}$ at
$z=(6.0, 7.0, 8.6)$, respectively, the red squares correspond to
$\rho_{\ast}=(0.27, 0.12, 0.05)\times 10^7$~$M_\odot$~Mpc$^{-3}$ at 
$z\approx (6.0, 7.0, 8.6)$, respectively, and the upper envelope 
corresponds to $(1.63, 1.37, 1.08)\times 10^7$~$M_\odot$~Mpc$^{-3}$ at
$z\approx (6.0, 7.0, 8.6)$, respectively. For comparison, the orange dotted line
here maps the orange dotted line in the top panel.
Note that the upper envelope of the GSMD at $z\approx 6$ is significantly
higher than what has actually been detected by {\it Spitzer} IRAC observations
(Yan et al. 2006; blue symbols).
This is not necessarily a discrepancy, because the latter
is a lower limit. However, it does suggest that the vast majority of the
stellar masses assembled during the reionization epoch are yet to be
detected at $z\approx 6$. Future observations with the JWST will be able to
determine whether this is the case.

\section{Summary}

   In this work, we have searched for galaxy candidates at $z\approx 7$ to 10
using the deepest ever near-IR observations obtained by the new WFC3 instrument
that was recently installed to the \hst. While these existing data are only 
from the first epoch of observations of the entire program, they have allowed
us to explore the universe at the highest redshifts ever possible. 

   By carefully reducing and analyzing these precious data, we are able to
take full advantage of the unprecedented depth that these new observations can
offer. Using the standard Lyman-break selection technique, we have found
20 $z_{850}$-dropouts, 15 $Y_{105}$-dropouts and 20 $J_{125}$-dropouts, which
are highly probable ($S/N>3$) candidate Lyman-break galaxies in three wide
redshift ranges at $z\approx 7$, 8 and 10, respectively. These are the largest
samples of very high-redshift galaxies to date. Among them, four
$z_{850}$-dropouts and ten $Y_{105}$-dropouts have not been reported by others,
and the objects in the entire $J_{125}$-dropout sample are new discoveries.
We have derived photometric redshifts for the $z_{850}$-dropouts by fitting
their multi-band SEDs, and the distribution of these photometric redshifts give
us an extra level of confidence that our dropout selection indeed selects
galaxies at high redshifts. While the $J_{125}$-dropouts are single-band
detections and hence are less secure as compared to the $z_{850}$- and
$Y_{105}$-dropouts, our test indicates that at least $>50$\% of them are very
likely genuine candidates at $z\approx 10$. We point out that the recent
criticism that the majority of our $J_{125}$-dropouts are implausibly too 
close to ``bright foreground" objects is not justified. While there are a few
cases that our dropouts are close to a neighbor, the excess fraction is 
$\sim$~30\%. While they could be due to some comtaminants of unknown origins,
we suggest that these objects could be genunine $z\approx 10$ galaxies that are
gravitationally lensed by their foreground neighbors, and that the seemingly
high rate could be explained by their intrinsically very steep LF and the
magnification bias. Future observations with the JWST will be able to prove or
refute this interpretation. We stress that including or excluding them from the
sample does not change our major conclusions.

   Our search for dropouts does not go beyond the limit that has reached by
other independent studies. While the number density of $z_{850}$-dropouts
agrees more or less with the expectations based on the previous works, 
the most surprising fact is that there are no bright candidates in the
$Y_{105}$- {\it and} $J_{125}$-dropout sample. The firm detections and the
firm non-detections above and below $\sim 28.0$~mag in {\it both cases} seem to
suggest a very steep increase of the surface density. While the current data
are not yet able to set stringent constraints because of the small number
statistics and the limited dynamic range in luminosity, we find that the sharp
increase could be explained by the exponential part of the Schechter function.
Motivated by this, we propose a rather unusual sets of Schechter function
parameters to describe the LFs at $z\approx 8$ and 10. As compared to their
counterpart LF at $z\approx 7$, our proposed LFs at $z\approx 8$--10 have
$M^*$ fainter by $\sim 2.0$~mag and $\Phi^*$ higher by a factor of 17--90. 
We caution that these LFs are still tentative and are not yet demanded by the
data. Nevertheless, they are allowed by the existing observations and agree
with the available data better than other alternatives. If these LFs still
hold at the level beyond our current detection limits, they
would imply that there is a sudden emergence of an extremely large number of
low-luminosity (by our local standard) galaxies when we look back in time to 
$z\approx 10$, and their persistence extends well into $z\approx 8$. While this
is totally unexpected, it is fully consistent with, and naturally fits in the
picture of the cosmic hydrogen reionization, which is believed to begin
at $z\approx 11$ and end at $z\approx 6$.  Such galaxies could
account for the entire ionizing photon budget at the reionization epoch; in
fact, it is likely that they would overproduce ionizing photons and
therefore either the escape fraction of Lyman photons must be extremely low
in galaxies that are below some certain luminosity threshold, or galaxies
below such threshold were not formed at all.

   Based on our dropout samples, we have derived the global SFR densities at
$z\approx 7$--10. The exact value of $\dot\rho_\ast$ depends on the exact form
of the LFs and also the limit down to which the integration of the Schechter
function is still valid. As the answers to both questions will remain highly
uncertain until the JWST is launched,
we derive our results based on a range of possibilities that cover
the most conservative estimate to the most radical one. If our proposed LFs at
$z\approx 8$ and 10 are indeed valid, they imply an extremely high global SFR
density ($\dot\rho_\ast$) in the early universe. 
Using $M=-15.0$~mag ($L=0.01\times L^*(z=3)$) as the fiducial limit, the 
integration of our LFs
show that $\dot\rho_\ast$ could start from 
$\sim 0.12$~$M_\odot$~yr$^{-1}$~Mpc$^{-3}$ at $z\approx 10$, 
rapidly decline to $\sim 0.01$~$M_\odot$~yr$^{-1}$~Mpc$^{-3}$ at $z\approx 7$,
and then start to rise again towards lower redshifts. 
A very high GSFRD at
$z\gtrsim 8$ is in line with the picture of the reionization, and has also been
suggested based on the study of long-duration GRB.
While the universe might have
started vigorously forming stars at $z\approx 10$ and seems have turned
$\sim 1.4\times 10^7$~$M_\odot$~Mpc$^{-3}$ worth of matter into stars over the
$\sim 300$~Myr to $z\approx 7$, the most massive galaxies at $z\approx 7$ are
still only to the order of a few $\times 10^9$~$M_\odot$. A large fraction
of stellar masses assembled during the reionization epoch seem undetected so
far at $z\approx 6$. The dramatic decrease of $\Phi^*$ from $z\approx 10$ to 7
probably suggests that big galaxies at $z\approx 7$--6 have gained their masses
mostly through merging of subsystems. We have detected a few cases of
close pairs and mergers indicative of such a scenario, however, the statistics
is still too limited to draw any meaningful constraint. This work, together
with those of other groups, show that it will be essential for the JWST to
fully explore the $z\gtrsim 8$ regime.

\acknowledgements

   We thank the WFC3 team for making this wonderful instrument to work and for
their timely delivery of the essential tools to make the reduction of new data
possible. We also acknowledge the team of the \hst\, Program GO-11563 for not
retaining a proprietary period of the data. We thank K. Chae, S. Mao, 
C. Kochanek, and D. Weinberg for valuable discussions. HY is supported by the
long-term fellowship program of the Center for Cosmology and AstroParticle
Physics (CCAPP) at The Ohio State University. RAW is supported in part by NASA
JWST Interdisciplinary Scientist grant NAG5-12460 from GSFC.

\end{document}